\documentclass[aps,pra,10pt,english,showpacs,floatfix,twocolumn,twoside]{revtex4-1}
\usepackage{amsmath,amstext,amssymb,babel,graphicx,geometry,bm,dcolumn,color,times}
\usepackage[colorlinks=true,citecolor=blue,urlcolor=red]{hyperref}
\usepackage[table]{xcolor}
\definecolor{red}{HTML}{000000}

\begin{document}

\title{Static and dynamical quantum correlations in phases of an alternating field XY model}
\author{Titas Chanda$^{1,2}$, Tamoghna Das$^{1,2}$, Debasis Sadhukhan$^{1,2}$, Amit Kumar Pal$^{1,2}$, Aditi Sen(De)$^{1,2}$, Ujjwal Sen$^{1,2}$}
\affiliation{$^1$Harish-Chandra Research Institute, Chhatnag Road, Jhunsi, Allahabad 211019, India \\
$^2$Homi Bhabha National Institute, Training School Complex, Anushaktinagar, Mumbai - 400094, India}

\begin{abstract}
We investigate the static and dynamical patterns of entanglement in an anisotropic XY model with an alternating transverse magnetic field, which is equivalent to a two-component one-dimensional Fermi gas on a lattice, a system realizable with current technology. Apart from the antiferromagnetic and paramagnetic phases, the model possesses a dimer phase which is not present in the transverse XY model. At zero temperature, we find that the first derivative of bipartite entanglement can detect all the three phases. We analytically show that the model has a ``factorization line" on the plane of system parameters, in which the  zero temperature state is separable. Along with investigating the effect of temperature on entanglement in a phase plane, we also report a non-monotonic behavior of entanglement with respect to temperature in the anti-ferromagnetic and paramagnetic phases, which is surprisingly absent in the dimer phase. Since the time dynamics of entanglement in a realizable physical system plays an important role in quantum information processing tasks, the evolutions of entanglement at small as well as large time are examined. Consideration of large time behavior of entanglement helps us to prove that in this model, entanglement is always ergodic. We observe that other quantum correlation measures  can  qualitatively show similar features in zero and finite temperatures. However, unlike nearest-neighbor entanglement,  the nearest-neighbor information theoretic measures can be both ergodic as well as non-ergodic, depending on the system parameters. 
\end{abstract}


\maketitle

\section{Introduction}
\label{intro}

Quantum many-body systems have been established to be a possible candidate for the implementation of quantum information protocols \cite{amico-rmp,lewenstein-rev} such as one-way quantum computation \cite{one-way-qc} and network quantum communication \cite{comun-spin}. Also, laboratory realization of model Hamiltonians in various substrates, including optical lattice \cite{coldatom,xxz-exp,optical-lattice}, ion traps \cite{lewenstein-rev,lab_xy_ion}, solid state systems \cite{solid_xy}, and NMR \cite{nmr}, have made possible the testing of properties of several information theoretic measures of quantum correlations, belonging to both of the entanglement-separability \cite{horodecki-rmp} and information-theoretic \cite{modi-rmp} domains. On the other hand, tools developed in, and with the help of, quantum information theory have been found to be useful in the analysis of the ground and excited states of such many-body systems \cite{sim-comp,dmrg-usual,dmrg-info}. Moreover, development of topological quantum computation and especially topological quantum memories indicate the importance of quantum many-body systems in the goal of practical realization of a quantum computer \cite{topo-q-comp}. Consequently, in recent years, characterization of quantum many-body systems from quantum information theoretic perspectives have become a vibrant field of research.

Although most of such studies are restricted to the ``static" properties of quantum correlations in the zero-temperature and thermal states, the time evolution of the system is also extremely important in quantum information 
processing tasks like in one way quantum computation \cite{one-way-qc}. In the static case, the traditional approach to study a quantum many-body system is to 
recognize appropriate order-parameters defining the 
phases occurring in the system, and to investigate the response of these order 
parameters to external perturbations. 
The ground state of such a system is usually represented by a complex multipartite quantum state, characterized by the classical 
as well as the quantum correlations present between its constituting parts. 
A quantum phase transition (QPT) \cite{qpt-books,dutta-qpt-books}, which occurs at zero temperature and solely due to 
quantum fluctuations, brings about a
qualitative change in the ground state of a quantum many-body system, when a system parameter is varied. Quantum correlations
having quantum information theoretic origins, 
are shown to be useful in characterizing 
various phases and corresponding QPTs in a large spectrum of quantum many-body systems 
\cite{XXZ-qpt,latorre-jphysa,eisert-jphysa,xy-qpt-ent,other-qpt-ent,other-nonmono,other-qpt-ent-2,syl,xy-qpt-disc,other-qpt-disc}
(see also \cite{amico-rmp,modi-rmp}, 
and the references therein).
Among all these models, a prominent one is the one-dimensional (1d) Fermi gas of spinless fermions in an 
optical lattice -- a system realizable in ultracold atom substrate, by using a Fermi-Bose mixture in the strong-coupling limit \cite{1dfermigas}. 
In the spin language, the model can be described by an anisotropic XY model in a transverse magnetic field 
\cite{lsm,bmpapers,qpt-books,dutta-qpt-books}.

Manipulation of cold atoms in the laboratory has allowed the realization of physical systems such as dilute atomic Fermi and Bose gases, 
in different spatial dimensions, thereby providing excellent opportunities to apply quantum information theoretic concepts in these 
systems \cite{bose-review,fermi-review}. 
Recent experimental evidences of superfluid, metallic, and Mott-insulating phases \cite{fermi-old-exp,fermi-recent-exp} motivates one to 
investigate a Fermi gas of spinless fermions in an 1d optical lattice, where the fermions are of two types, 
distinguished by different chemical potentials. Considering the two types of fermions to be located  on two different sublattices, 
one of which contains all the  ``even'' sites and the other one holds all the ``odd'' ones, 
the fermionic model, via a Jordan-Wigner transformation, can be shown to be equivalent to an 1d anisotropic XY model in the presence 
of a uniform, and an alternating transverse magnetic field that alternates its direction from $+z$ to $-z$
depending on whether the lattice site is even, or odd \cite{dutta-qpt-books,kogut,alt-field-old,alt-field-defect,diep}. 
The model offers a rich phase diagram. While only two phases, viz. a  ``paramagnetic'' (PM) phase and an 
``antiferromagnetic'' (AFM) phase occur in the ground state of the XY model in a uniform transverse field, \cite{qpt-books,bmpapers,lsm}, 
an additional ``dimer'' (DM) phase emerges due to the introduction of the local site-dependent alternating field in the present 
model \cite{dutta-qpt-books,kogut,alt-field-old,alt-field-defect,diep}. Although the properties of several quantum information theoretic 
measures of quantum correlations have been extensively studied and reported for different phases and corresponding QPTs in the former 
case \cite{xy-qpt-ent,xy-qpt-disc,syl} (see also references in \cite{amico-rmp,modi-rmp}), 
it is interesting to see how the new phase structure, formed due to the introduction of the alternating field, can be characterized 
using quantum correlations. In this paper, we characterize the static as well as dynamic properties of quantum correlations in the 
1d anisotropic XY model in a uniform and an alternating field. 
As the measures of quantum correlations, we focus on bipartite measures, and use logarithmic negativity (LN) \cite{neg_group} from the 
entanglement-separability genre, and quantum discord (QD) \cite{disc_group,total_corr} from the information-theoretic domain.
We show that irrespective of the values of the anisotropy parameter, first 
derivative of bipartite entanglement can detect all the three phases in this model. Moreover, the finite-size scaling analysis
of the system near the QPTs is performed to distinguish phase boundaries between the AFM and the PM, and between the AFM and the DM. 
Similar investigations are also carried out for quantum discord, which also faithfully indicate the quantum critical points. 
Like the factorization point in the XY model \cite{factor_old,factor_new}, 
we here prove 
 the existence of a line in the space of the system parameters, which we call as the ``factorization line'' (FL), on which the ground 
state of the system is separable, having a N\'{e}el-type order. 

The change of phase diagram with finite temperature has both fundamental and experimental importance due to the technological limitations of reaching absolute zero temperature. 
In this scenario, we discuss the weathering of the landscapes of quantum correlations over the phase-plane of the system 
parameters, chosen to be the strengths of the uniform and the alternating transverse field, with increasing 
temperature. We point out that bipartite entanglement is the most fragile in the AFM phase, while it is robust in the DM phase against 
increasing temperature. 
 We identify the phases in which nonmonotonicity of entanglement with the increase of temperature is observed. 
Specifically, we perform a non-monotonicity cartography, and map, on the plane of the chosen system parameters, the regions in which 
the thermal quantum correlations exhibit non-monotonic variation with temperature. We show that for LN and for high values of anisotropy 
parameter, most of the non-monotonicity occurs in the AFM region, while QD is found to be nonmonotonic in the PM phase for low anisotropy. 
 Interestingly, we discover that the temperature variation of LN is found to be monotonic in the entire DM phase, while for QD, 
non-monotonicity occurs at a very small region of the DM phase.

As already stated, the time dynamics of quantum correlations in any physical system is extremely relevant for implementation of quantum information 
processing tasks. In this paper, we find both the small and large time quantum 
correlation patterns of the evolved state. We observe that although entanglement 
dies quickly compared to QD, it possesses larger value than QD, which 
ensures the possibility of implementing several information tasks requiring 
high values of entanglement. The study of large time behaviour of quantum 
correlations also helps us to settle issues like 
the ergodicity \cite{ergodic_def,dyn-ergo,dyn-ergo-2,dyn-ergo-3,dyn-ergo-xy,dyn-ergo-xy-dual,dyn-ergo-xy-sen,dyn-ergo-xy-group,dyn-ergo-other} of LN and QD, quantified by the ergodicity scores. We find that, up to our numerical accuracy, entanglement always remains ergodic, while QD shows nonergodicity 
in different phases of the model. We point out that the region of nonergodicity of QD increases with an increase in the anisotropy in the 
system. 
Therefore with respect to transverse field parameter, we show that QD undergoes a nonergodic to ergodic transition which is absent for entanglement upto our numerical accuracy, irrespective of the anisotropy parameter and initial temperature.

The paper is organized as follows. In Sec. \ref{model}, the Hamiltonian describing the 
anisotropic XY model in the presence of a uniform,  and an alternating transverse field, and its relation to
a two-component 1d Fermi gas are discussed. 
Brief descriptions on the diagonalization of the model Hamiltonian, and the different phases occurring in the ground state of the model 
are provided in the same section. Sec. \ref{two-qubit} contains the definitions of the canonical equilibrium state and the time-evolved state
of the system. The determination of the single-site and two-site reduced density matrices from the canonical equilibrium state and the 
time-evolved state of the model is also presented in this section. The static properties of the quantum correlations, including the 
different types of QPTs, finite-size scaling analysis, determination of the factorization line, and thermal 
quantum correlations are discussed in Sec. \ref{static_qc}. Sec. \ref{dynamic} reports the ergodicity of quantum correlations 
and short-time dynamics of entanglement as well as QD. Sec. \ref{conclude} contains the concluding remarks.

\section{The model}
\label{model}

Let us consider a family of models describing a system of spins of magnitude $\frac 12$ on an 1d
lattice consisting of $N$ sites. 
We assume that an external transverse magnetic field of site-dependent strength $h_i(t)=h_1(t)+(-1)^ih_2(t)$, $i$ being the 
site index, acts on the spins at time $t$. The magnetic field can be interpreted as the resultant of a uniform transverse field, $h_1(t)$, 
and a transverse field, $h_2(t)$, which reverses its direction from $+z$ to $-z$, depending on whether the lattice site is even, or odd. 
The Hamiltonian describing the system is given by 
\begin{eqnarray}
\hat{H}&=&\frac{1}{2}\sum_{i = 1}^{N}\Big[J\Big\{\frac{1+\gamma}{2}\hat{\sigma}_i^x\hat{\sigma}_{i+1}^{x}+\frac{1-\gamma}{2}\hat{\sigma}_i^y\hat{\sigma}_{i+1}^{y}\Big\}\nonumber \\
&&+(h_1(t)+(-1)^ih_2(t))\hat{\sigma}_i^z\Big].
\label{ham_spin}
\end{eqnarray}
Here, the system parameter $J$ represents the strength of the exchange interaction, while $\gamma (\neq 0)$ is the $x-y$ anisotropy 
present in the system. We assume periodic boundary condition (PBC), and an even number of lattice sites, such that 
$\hat{\sigma}^{\alpha}_{N+1}\equiv \hat{\sigma}_1^\alpha$, where $\alpha=x,y,z$.

\subsection{Relation to one-dimensional Fermi gas}
\label{fermion_model}

The Hamiltonian in Eq. (\ref{ham_spin}), via a Jordan-Wigner transformation, given by \cite{alt-field-defect}
\begin{eqnarray}
\hat{\sigma_{2j}}^{+}&=&\hat{b}_{2j}^{\dagger}\exp{\Big(i\pi\sum_{l=1}^{i-1}
\hat{b}_{2l}^{\dagger}\hat{b}_{2l}+i\pi \sum_{l=1}^{i}\hat{a}_{2l-1}^{\dagger}\hat{a}_{2l-1}\Big)},
\nonumber \\
\hat{\sigma}_{2j+1}^{+}&=&\hat{a}_{2j+1}^{\dagger}\exp{\Big(i\pi\sum_{l=1}^{i}
\hat{b}_{2l}^{\dagger}\hat{b}_{2l}+i\pi \sum_{l=0}^{i-1}\hat{a}_{2l+1}^{\dagger}\hat{a}_{2l+1}\Big)},\nonumber \\
\label{jwt}
\end{eqnarray}
can be mapped onto a two-component Fermi gas of spinless fermions, on an 1d optical lattice consisting of two sublattices.
Here, $\hat{\sigma}^-_{\alpha}=(\hat{\sigma}^+_\alpha)^\dagger$, where the $\hat{\sigma}_\alpha^{\pm}$ operators are related to the 
Pauli operators $\hat{\sigma}^{x,y,z}$ via the relations $\hat{\sigma}^x=(\hat{\sigma}^+ +\hat{\sigma}^-)$, 
$\hat{\sigma}^y=-i(\hat{\sigma}^+-\hat{\sigma}^-)$, and $\hat{\sigma}^z=(2\hat{\sigma}^+\hat{\sigma}^--1)$.
One of the two sublattices in the fermionic model is constituted of the ``odd'' lattice sites, while the other contains the ``even'' ones. 
One of the two components of the fermions is situated on the odd sublattice, while the 
other is located on the even sublattice. The two components are distinguished by two different time-dependent 
chemical potentials, $\mu_a(t)$ and $\mu_b(t)$, 
and the corresponding creation operators are denoted by $\hat{a}^\dagger$ and $\hat{b}^\dagger$, respectively, following the usual fermionic 
anticommutation relations $\{\hat{f}_i,\hat{f}_j^\dagger\}=\delta_{i,j}$, and $\{\hat{f}_i,\hat{f}_j\}=\{\hat{f}_i^\dagger,\hat{f}_j^\dagger\}=0$. Here, $\hat{f}=\hat{a}$ or 
$\hat{b}$, depending on whether $i,j$, the site indices, 
are odd or even, respectively.

Applying the transformation in Eq. (\ref{jwt}), the form of the 
Hamiltonian representing the 1d two-component Fermi gas of spinless fermions
at every time instant $t$, up to an additive constant energy $E_c(t)=(\mu_a(t)+\mu_b(t))N/4$, can be written as 
\begin{eqnarray}
\hat{H} &=&\sum_{i=1}^{N/2}\Big[\tau\big\{\hat{\mathcal{A}}_i+\hat{\mathcal{B}}_i
+\gamma(\hat{\mathcal{C}}_i+\hat{\mathcal{D}}_i)\big\}
+\mu_a(t)\hat{\mathcal{N}}^a_i+\mu_b(t)\hat{\mathcal{N}}^b_i\Big],\nonumber\\
\label{ham_ferm}
\end{eqnarray}
where the operators $\hat{\mathcal{A}}_i=\hat{a}_{2i-1}^{\dagger}\hat{b}_{2i}+h.c.$, 
$\hat{\mathcal{B}}_i=\hat{b}_{2i}^{\dagger}\hat{a}_{2i+1}+h.c.$, 
$\hat{\mathcal{C}}_i=\hat{a}_{2i-1}^{\dagger}\hat{b}_{2i}^{\dagger}+h.c.$, and 
$\hat{\mathcal{D}}_i=\hat{b}_{2i}^{\dagger}\hat{a}_{2i+1}^{\dagger}+h.c.$ describe the interactions 
between the spinless fermions belonging to  the odd and the even sublattices, 
with $\hat{\mathcal{N}}^a_i=\hat{a}_{2i-1}^{\dagger}\hat{a}_{2i-1}$ and 
$\hat{\mathcal{N}}^b_i=\hat{b}_{2i}^{\dagger}\hat{b}_{2i}$ being the corresponding number operators. Here, $\tau$ is the fermionic tunneling 
strength between a pair of even and odd sites, 
and $N$ is the total number of lattice sites. 
Note that the existence of the two types of magnetic field (uniform and alternating) in the original model is reflected by the existence 
of the two sublattices in the fermionic model, differentiated by the chemical potentials and thereby leading to
two types of fermionic operators, $a$ and $b$.

\subsection{Diagonalization}
\label{diagonalization}

For general $\mu_{a,b}(t)$, the Hamiltonian given in Eq. (\ref{ham_ferm}) can be written
as $\hat{H}=\sum_{p=1}^{N/4}\hat{H}_p$, with 
\begin{eqnarray}
 \hat{H}_p &=& J\cos{\phi_p}(\hat{a}_p^{\dagger}\hat{b}_p+a_{-p}^{\dagger}
 \hat{b}_{-p}+\hat{b}_p^{\dagger}\hat{a}_p+\hat{b}_{-p}^{\dagger}\hat{a}_{-p})\nonumber \\
 &&-iJ\gamma\sin{\phi_p}(\hat{a}_p^{\dagger}\hat{b}_{-p}^{\dagger}+\hat{a}_{p}b_{-p}-\hat{a}_{-p}^{\dagger}\hat{b}_p^{\dagger}-\hat{a}_{-p}a_{p})\nonumber\\
 &&+h_+(t)(\hat{b}_p^{\dagger}\hat{b}_p+\hat{b}_{-p}^{\dagger}\hat{b}_{-p})
 +h_-(t)(\hat{a}_p^{\dagger}\hat{a}_p+\hat{a}_{-p}^{\dagger}\hat{a}_{-p})\nonumber\\
 &&-2h_1(t)\nonumber \\
 \label{hp}
\end{eqnarray}
via the Fourier transformations given by  
\begin{eqnarray}
 \hat{a}_{2j+1}^{\dagger}&=&\sqrt{\frac{2}{N}}\sum_{p=-N/4}^{N/4}\exp{\big(i(2j+1)\phi_p \big)}\hat{a}_p^{\dagger}, \nonumber \\
 \hat{b}_{2j}^{\dagger} &=&\sqrt{\frac{2}{N}}\sum_{p=-N/4}^{N/4} \exp{\big(i(2j)\phi_p \big)} \hat{b}_p^{\dagger}.
\end{eqnarray}
Here $\phi_p=2\pi p/N$, $h_\pm(t)=h_1(t)\pm h_2(t)$, and $a_p^\dagger$ ($b_p^\dagger$) is fermionic operators. 
Since $[\hat{H}_p,\hat{H}_{p^\prime}]=0$, the above Fourier transformation decomposes the space upon which $\hat{H}$ acts into non-interacting 
subspaces. These subspaces, each having a dimension sixteen, do not allow transitions within themselves, irrespective of the values 
of the system parameters $J$, $\gamma$, and $h_\pm(t)$. The diagonalization of the Hamiltonian $\hat{H}$ is thereby reduced to the diagonalization 
of $\hat{H}_p$, acting on the $p^{th}$ subspace, which can be achieved by a convenient choice of the basis 
(see Appendix \ref{ap:diagonalization}). We note that the \textcolor{red}{lowest eigenvalue  of $\hat{H}_p$, is given by $-\omega^4_+(p)$. The ground state energy per site, $E_0$, of the Hamiltonian can be obtained as 
$E_0=-\frac{1}{2\pi}\int_0^{\pi/2}\omega^4_+(p)dp$. }

\begin{figure*}
 \includegraphics[width=\textwidth]{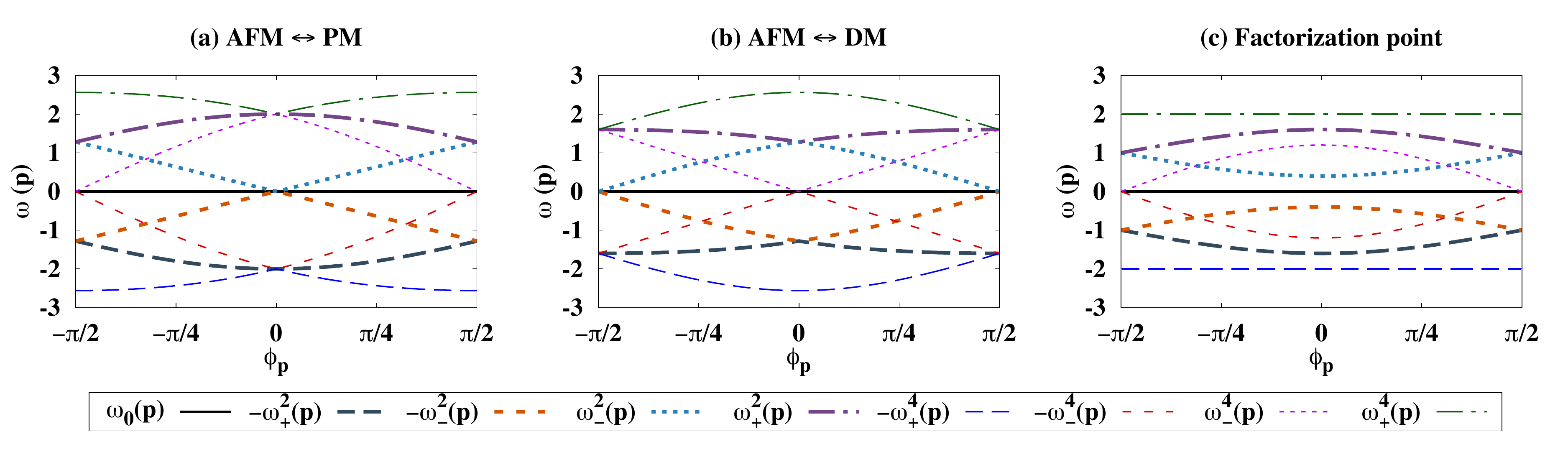}
 \caption{\textcolor{red}{(Color online.) Spectrum of $\hat{H}_p$ for $\gamma=0.8$.  \textbf{(a), (b)} Variation of the eigenvalues of $\hat{H}_p$ as a function of $\phi_p$ at the AFM $\leftrightarrow$ PM ($\lambda_1=1, \lambda_2=0$),  and the AFM $\leftrightarrow$ DM  ($\lambda_1=0, \lambda_2=0.8$) transitions. \textbf{(c)} Patterns of the eigenvalues of $\hat{H}_p$ against $\phi_p$ at the factorization point (see Sec. \ref{stat:factor}). The chosen parameter values for the factorization point are $\lambda_1=0.6, \lambda_2=0$.  The minimum eigenvalue in all the cases is $-\omega^4_+(p)$ (given in Appendix \ref{ap:diagonalization}). The energy gap vanishes at $\phi_p = 0$ for the AFM $\leftrightarrow$ PM, and at $\phi_p = \pm \frac{\pi}{2}$ for the AFM $\leftrightarrow$ DM QPT point.}}
 \label{spectrum}
\end{figure*}

\subsection{Phases}
We now briefly discuss the patterns of different phases, and the corresponding QPTs, present in the model described by the Hamiltonian in Eq. (\ref{ham_ferm}). We choose the strength of the transverse fields, uniform and alternating, as the tuning parameters. 
Information about the phase-boundaries can be obtained from the second-order derivatives of the ground state energy, $E_0$, with respect to $\lambda_1$ and $\lambda_2$, where we take $\lambda_i = h_i/J, ~~ i = 1,2$ and $h_{1(2)}(t=0)=h_{1(2)}$. For $\gamma \neq 0$, the system undergoes two different second-order QPTs, namely, a transition from a paramagnetic (PM) to an antiferromagnetic (AFM) phase, and a transition from the AFM to a dimer (DM) phase. \textcolor{red}{Fig.\ref{spectrum}(a) and (b) depict the spectrum of $\hat{H}_p$ at critical points corresponding to AFM $\leftrightarrow$ PM ($\lambda_1=1, \lambda_2=0$), and AFM $\leftrightarrow$ DM ($\lambda_1=0, \lambda_2=0.8$) QPTs, respectively, for $\gamma=0.8$ (see Appendix \ref{ap:diagonalization} for the expressions of the eigenvalues as functions of $\phi_p$ and the system parameters). Note that the vanishing of the energy gap in the spectrum occurs at $\phi_p = 0$ for the AFM $\leftrightarrow$ PM transition, and at $\phi_p = \pm \frac{\pi}{2}$ for AFM $\leftrightarrow$ DM transition. On the other hand,  Fig. \ref{spectrum}(c) depicts the variation of the spectrum of $\hat{H}_p$ as a function of $\phi_p$ for $\lambda_1=0.6, \lambda_2=0$. This point on the $(\lambda_1,\lambda_2)$ plane belongs to the \emph{factorization line}, which  is discussed in Sec. \ref{stat:factor}.} One of our aims in this paper is to detect such transitions by using quantum information quantities. The phase boundaries corresponding to these transitions are given by the lines $\lambda_1^2=\lambda_2^2+1$, and $\lambda_2^2=\lambda_1^2+\gamma^2$, respectively. It is interesting to note that there exists a set of duality relations, given by $\{h_1\leftrightarrow h_2, J\leftrightarrow -\gamma\}$, by virtue of the unitary transformation $\{\hat{\sigma}_i^\alpha\rightarrow(-1)^i\hat{\sigma}_i^\alpha: \alpha=x,z\}$, which indicates that both AFM $\leftrightarrow$ PM and AFM $\leftrightarrow$ DM transitions belong to the same universality class, namely, the Ising universality class \cite{dutta-qpt-books}. One must also note that for $h_2=0$, the model reduces to the well-known anisotropic XY model in a uniform transverse magnetic field of magnitude $h_1$.

\section{Canonical-equilibrium and time-evolved states: Local density matrices}
\label{two-qubit}

In this paper, we intend to study the statistical mechanical properties of the model in terms of bipartite quantum correlations. 
We now briefly introduce the notions of canonical equilibrium states and time-evolved states corresponding to the Hamiltonian 
given in Eq. (\ref{ham_spin}), and describe how two-spin reduced density matrices corresponding to such states can be obtained. 
For our purpose,  we consider the situation where the time-dependent magnetic fields $h_1(t)$ and $h_2(t)$, are chosen as
\begin{eqnarray}
 h_1(t)= \left\{
 \begin{array}{cc}
 h_1, & t\leq 0  \\
 0, & t>0
\end{array}\right.,\;\;
 h_2(t)= \left\{
 \begin{array}{cc}
 h_2, & t\leq 0  \\
 0, & t>0
\end{array}\right..
\end{eqnarray}

The canonical equilibrium state (CES) of the system at time $t$ is given by 
\begin{eqnarray}
\hat{\rho}_{eq}(t)=\frac{e^{-\beta \hat{H}(t)}}{Z},
\label{ces}
\end{eqnarray}
where $Z=\mbox{Tr}\left[\exp(-\beta \hat{H}(t))\right]$ is the partition function,
and $\hat{H}(t)$ is given in Eq. (\ref{ham_spin}). Here, $\beta=1/k_{B}T$, $T$ is the 
absolute temperature, and $k_B$ is the Boltzmann constant.  In all our calculations, we set $k_B=1$. 
For the purpose of this paper, we consider a system which is in contact with a heat bath at temperature $T$ for a long time up to the instant 
that we call $t=0$, so that a thermal equilibrium between the system and the heat bath have developed. The equilibrium is in the 
canonical sense, allowing exchange of energy between the bath and the system with the usual average energy constraint, but forbidding 
exchange of particle. To study quantum correlations in the evolution, we choose the canonical equilibrium state $\big(\hat{\rho}_{eq}(t = 0) \big)$ as an initial state. When the magnetic fields are switched off, the CES starts evolving in time following the Schr\"{o}dinger equation
dictated by the Hamiltonian in Eq. (\ref{ham_spin}). 
At any time $t$, the time-evolved state (TES), $\hat{\rho}(t)$, is given by 
\begin{eqnarray}
 \hat{\rho}(t)=e^{-i\hat{H}t}\hat{\rho}_{eq}(t=0)e^{i\hat{H}t},
 \label{tes}
\end{eqnarray}
where $\hat{H}$ represents the Hamiltonian given in Eq. (\ref{ham_spin}) at $t>0$. 

\subsection{Local density matrices}
\label{ldm}

To investigate the behaviour of bipartite quantum correlation measures of the CES and TES,  
computation of the single-site and the two-site reduced density
matrices of the entire state is necessary. Since we consider the system with periodic boundary condition, 
all the nearest neighbour bipartite state are same and hence their
two-spin correlation functions would be independent of the choice of the pairs of spins, while the single-site magnetizations depends 
on whether the lattice site is even, or odd.
A general single-site density matrix, given by $\hat{\rho}^i=[\mathbb{I}+\sum_{\alpha=x,y,z}m^\alpha(t)\hat{\sigma}^\alpha_i]/2$,
can be obtained by tracing out all the spins except the spin 
at the lattice site $\alpha$, which is ``$o$'' for the odd site, and ``$e$'' for the even site. Here, 
$\mathbb{I}$ is the identity operator in the qubit Hilbert space. 
For CES corresponding to a real Hamiltonian, 
$\hat{\rho}_{eq}^{i*}(t)=\hat{\rho}_{eq}^{i}(t)$, implying $m^y_{i}(t)=0$ with 
the complex conjugation being taken in the computational basis. 
Also, the Hamiltonian possesses a global phase-flip symmetry, 
such that $[H,\Pi_i\sigma_i^z]=0$, implying $m^x_{i}(t)=0$. Hence, the single-site reduced density matrix corresponding to the CES 
is given by $\hat{\rho}^i_{eq}(t)=(\mathbb{I}+m^z_i(t)\hat{\sigma}_i^z)/2$. On the other hand, $\hat{\rho}^{i}(t)$ corresponding to the 
evolved state is not necessarily equal to its complex conjugation, and the existence of the global phase-flip symmetry is a 
complicated issue due to the time dependence of the Hamiltonian. However, use of the Wick's theorem leads to the same form of 
$\hat{\rho}^i(t)$, when TES is considered instead of the CES. 

Let us now  consider the two-site reduced density matrix $\hat{\rho}_{ij}$, corresponding to the spins at the lattice sites 
$i$ and $j$, and obtained by tracing out all the other spins except those at the positions $i$ and $j$. 
In the present case, we restrict ourselves to nearest-neighbor
pairs of spins, such that $j\equiv i+1$. To keep the notations uncluttered, from now on, we shall discard the 
lattice indices, and denote the nearest neighbor two-spin density matrix by $\hat{\rho}_{eo}$, where we assume that the lattice site $i$ belongs 
to the even sublattice without any loss of generality.  
The two-party state $\hat{\rho}_{eo}$, of the CES and TES in this system, can be written as 
\begin{eqnarray}
 \hat{\rho}_{eo}&=&\frac{1}{4}\Big[\mathbb{I}_e\otimes\mathbb{I}_o
 +m^z_e \hat{\sigma}^z_e\otimes\mathbb{I}_o+\mathbb{I}_e\otimes m^z_o \hat{\sigma}^z_o\nonumber\\
 &&+\sum_{\alpha,\beta=x,y,z}c^{\alpha\beta}_{eo}\hat{\sigma}^\alpha_e\otimes\hat{\sigma}^\beta_o\Big],
\label{rhoij}
\end{eqnarray}
where $c_{eo}^{\alpha\beta}=\mbox{Tr}[\hat{\sigma}_e^\alpha\otimes\hat{\sigma}_o^\beta\hat{\rho}_{eo}]$ are the two-site 
spin correlation tensor. 
In the case of CES, by using arguments similar to that in the case of the single-site 
density matrix, and by applying the Wick's theorem,
one can show that only diagonal elements of the correlation tensor, given by $c^{\alpha\alpha}_{eo}$, $\alpha=x,y,z$, remain. 
On the other hand, in the case of TES, $c^{xy}_{eo}$ and $c^{yx}_{eo}$ remain non-zero in addition to the diagonal correlators.
For brevity, from now onward, we discard the site indices while mentioning the two-spin correlators.

\begin{figure}
 \includegraphics[scale=0.3175]{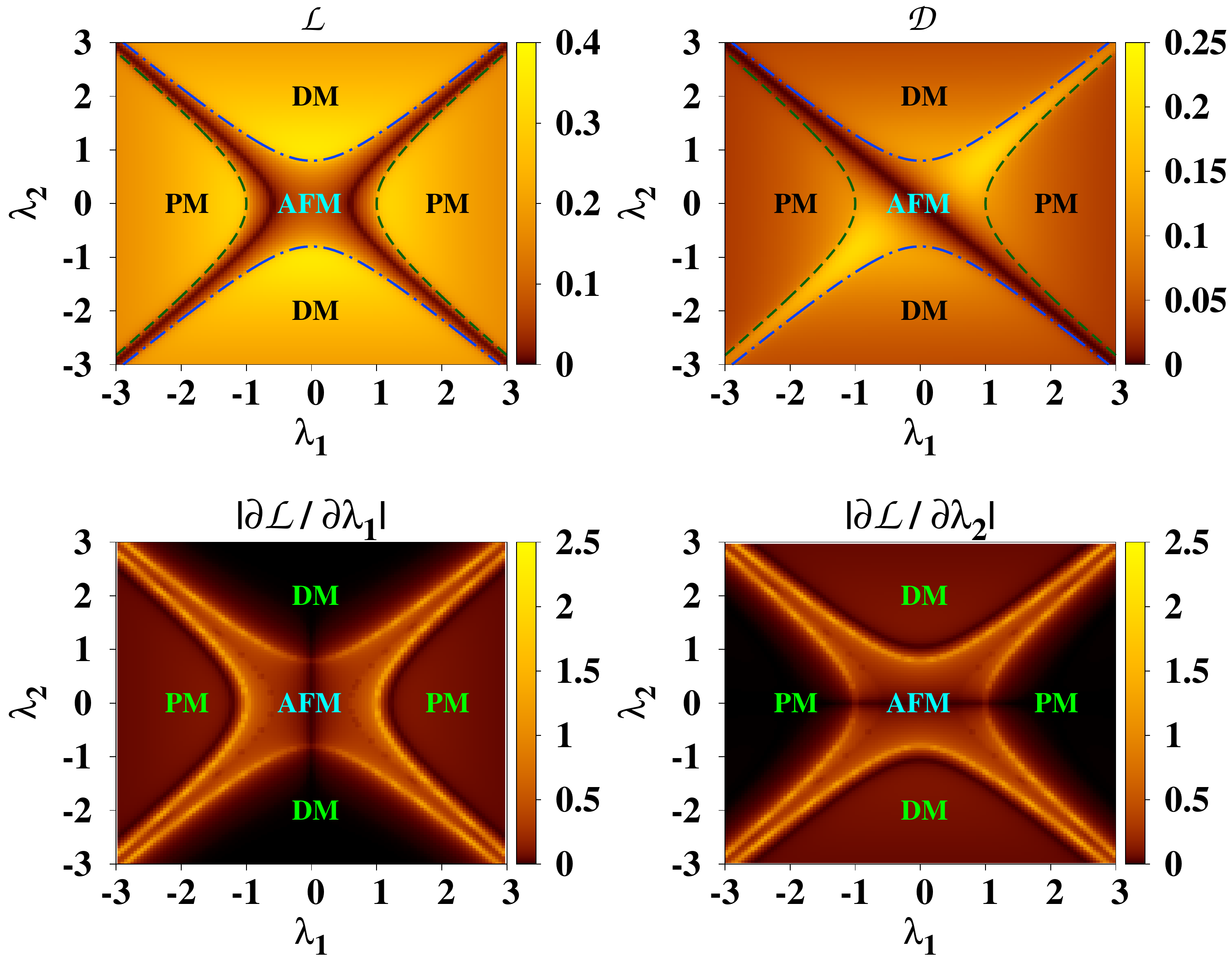}
 \caption{(Color online) \textbf{(Top horizontal pannels)} Variations of LN (left pannel) and QD (right pannel) as functions of the 
 transverse magnetic field $\lambda_1$ and the alternating field $\lambda_2$
 in the thermodynamic limit at $\beta\rightarrow\infty$, and $\gamma=0.8$. 
 The phase boundaries $\lambda_1^2=\lambda_2^2+1$ (PM $\leftrightarrow$ AFM) and 
 $\lambda_2^2=\lambda_1^2+\gamma^2$ (AFM $\leftrightarrow$ DM) are represented by the dashed and dot-dashed lines, respectively, while the 
 different shades in the figures represent different values of quantum correlations.
 \textbf{(Bottom horizontal pannels)} Variations of the first derivative of LN with respect to $\lambda_1$ (left pannel), 
 and the same quantity of LN with respect to $\lambda_2$ (right pannel)
with $N\rightarrow\infty$ at $\beta\rightarrow\infty$, and $\gamma=0.8$. The value of the respective first derivatives of LN diverges at   
 the phase boundaries $\lambda_1^2=\lambda_2^2+1$ (PM $\leftrightarrow$ AFM) and 
 $\lambda_2^2=\lambda_1^2+\gamma^2$ (AFM $\leftrightarrow$ DM). 
 Different shades in the figures represent different values of the first derivative of LN with respect to respective parameter. 
 All the quantities plotted in all the figures are dimensionless, except LN which is in ebits and QD in bits.}
 \label{lnqdggm}
\end{figure}

\subsection{Quantum correlations between two modes of a 1d Fermi gas}
\label{equiv}

We now demonstrate that the quantum correlation between a nearest-neighbour spin pair chosen from the anisotropic XY model in 
a uniform and an alternating transverse magnetic field is the same as that present between two fermionic modes located at the two 
nearest-neighbour lattice sites in the fermionic model given in Eq. (\ref{ham_ferm}). Without any loss of generality, the two-site 
density matrix of a nearest-neighbour pair of lattice sites, denoted by ``\textit{eo}'', can be written as 
$\hat{\rho}^f_{eo}=\frac{1}{4}\sum_{k,l}\xi_{kl}\hat{\varsigma}_e^k\hat{\varsigma}_o^l$,
where $k,l=0,1,2,3$, and $\hat{\varsigma}_{\alpha}=\{\mathbb{I},(c_\alpha+c_\alpha^\dagger),-i(c_\alpha-c_\alpha^\dagger),
(2c_\alpha^\dagger c_\alpha-1)\}$. Here, $c\equiv a(b)$ depending on whether $\alpha\equiv o(e)$. The coefficients, $\{\xi_{kl}\}$, 
are given by $\xi_{kl}=\mbox{tr}[\hat{\rho}_{eo}^f(\hat{\varsigma}_e^k\hat{\varsigma}_o^l)^\dagger]$. 
Expanding and applying Wick's theorem as in Sec. \ref{ldm}, the TES corresponding to a pair of fermionic modes on the ``\textit{eo}'' 
site pair for the fermionic model is given by 
\begin{eqnarray}
 \hat{\rho}^f_{eo}&=&\frac{1}{4}\Big[\mathbb{I}_{4\times 4}+\xi_{03}\hat{\varsigma}^3_{o}+\xi_{30}\hat{\varsigma}^3_{e}
 +\xi_{11}\hat{\varsigma}_e^1\hat{\varsigma}_{o}^1
+\xi_{22}\hat{\varsigma}^2_{e}\hat{\varsigma}^2_{o}+\xi_{33}\hat{\varsigma}^3_{e}\hat{\varsigma}^3_o\nonumber\\
&&+\xi_{12}\hat{\varsigma}^1_e\hat{\varsigma}^2_o
+\xi_{21}\hat{\varsigma}^2_{e}\hat{\varsigma}^1_{o}\Big].
\label{rho_expand}
\end{eqnarray}
With a convenient choice of basis given by $\{|0\rangle,\hat{a}^\dagger|0\rangle,\hat{b}^\dagger|0\rangle,
\hat{b}^\dagger\hat{a}^\dagger|0\rangle\}$, where $|0\rangle$ represents the vaccume state, the individual terms in Eq. (\ref{rho_expand})
can be expressed in their respective matrix forms. A comparison with the matrix forms of the 
operators $\sigma_e^\alpha\otimes\sigma_o^\beta$ implies that in matrix form, $\hat{\rho}^f_{eo}$ can be expressed 
as
\begin{eqnarray}
 \hat{\rho}^f_{eo}&=&\frac{1}{4}\Big[\mathbb{I}_e\otimes\mathbb{I}_o
 -m^z_e\sigma^z_e\otimes\mathbb{I}_o-\mathbb{I}_e\otimes m^z_o\sigma^z_o
 -c_{eo}^{xy}\sigma_e^x\otimes\sigma_o^y\nonumber\\
 &&-c_{eo}^{yx}\sigma_e^y\otimes\sigma_o^x
 +\sum_{\alpha=x,y,z}c^{\alpha\alpha}_{eo}\sigma^\alpha_e\otimes\sigma^\beta_o
 \Big].
\label{rho_fermi_equiv} 
\end{eqnarray}
Here, $\sigma^\alpha$ are $2\times 2$ the Pauli matrices, where e.g. $\sigma_z^e = \scriptsize \left(\begin{array}{c c}
1 & 0 \\
0 & -1\\
\end{array}\right)\normalsize$ in the $\{ |0\rangle, b^{\dagger}|0\rangle\}$ basis and where e.g. $\sigma_y^o = \scriptsize \left(\begin{array}{c c}
0 & -i \\
i & 0\\
\end{array}\right)\normalsize$ in the $\{ |0\rangle, a^{\dagger}|0\rangle\}$ basis.
Note that $\rho^f_{eo}$ is connected to the TES $\hat{\rho}_{eo}$ in the spin model via a local unitary transformation given by 
$\rho^f_{eo}=(\sigma^x\otimes\sigma^x)\rho_{eo}(\sigma^x\otimes\sigma^x)$, thereby implying no change in the values of the chosen measure of 
bipartite quantum correlation.  

\section{Static behaviour of quantum correlations}
\label{static_qc}

In this section, we discuss the behaviour of bipartite quantum correlation measures of 
the reduced density matrix of the nearest-neighbour qubit pair, obtained from the zero-temperature  and the thermal states of the model. 
Since the model is not evolving, we call the states as static states. 
For our purpose, we consider logarithmic negativity (LN), \textcolor{red}{denoted by $\mathcal{L}(\rho_{AB})$}, and 
quantum discord (QD), \textcolor{red}{denoted by $\mathcal{D}(\rho_{AB})$}, in the ground and thermal states of the model. 
The former belong to the entanglement separability paradigm, 
while the latter is from the quantum information theoretic regime of quantum correlations.
Short descriptions of these measures are provided in Appendix \ref{ap:qc}. While computing QD in the entire paper,
we always perform local rank-$1$ projection measurement on the ``even'' qubit.
We choose two different types of quantum correlation quantities since they behave differently as demonstrated in the XY as well as XXZ model \cite{xy-qpt-ent,xy-qpt-disc, XXZ-qpt}. 

\begin{figure*}
 \includegraphics[width=0.9\textwidth]{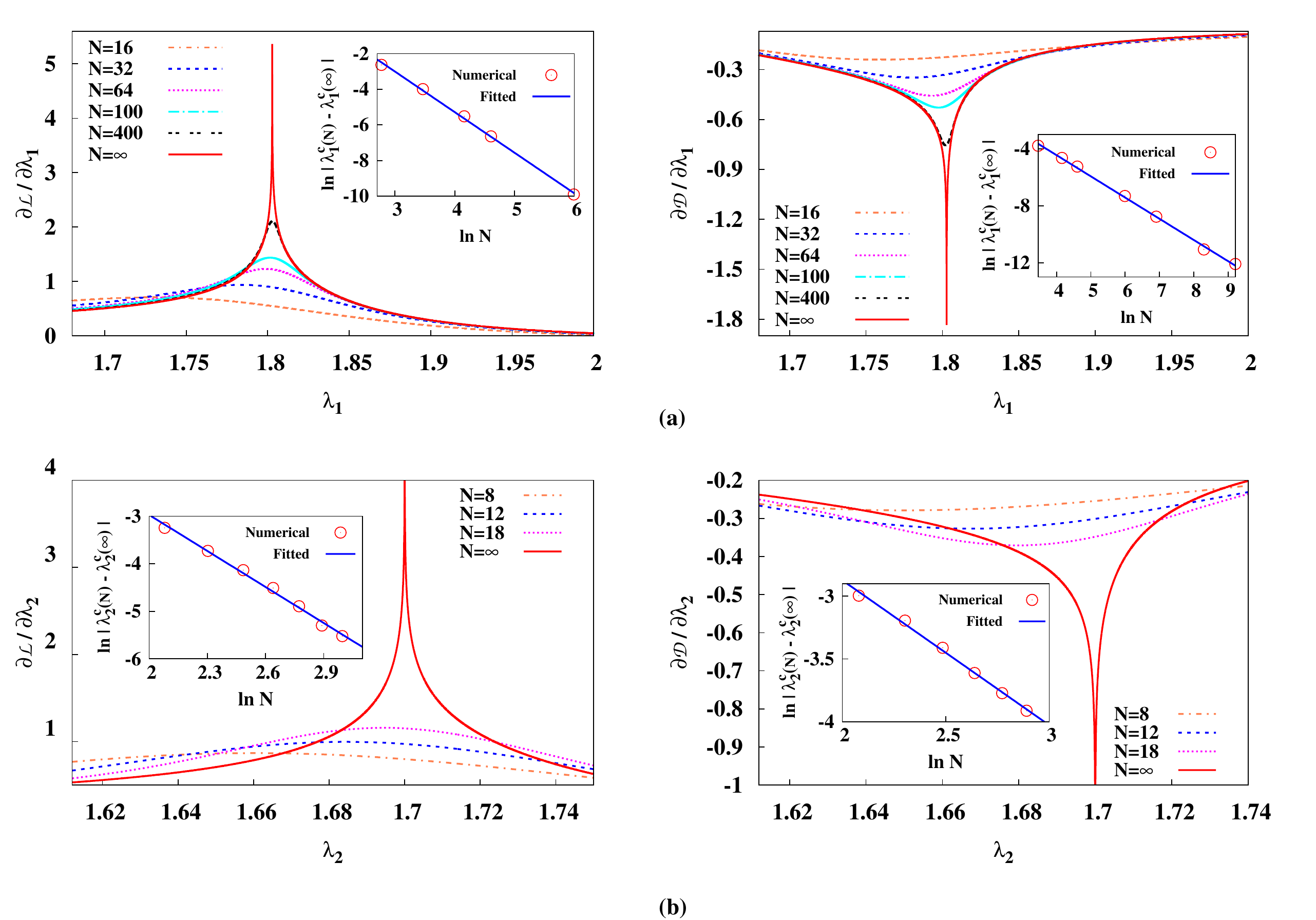}
 \caption{(Color online)  Finite-size scaling using LN and QD in (a) AFM $\leftrightarrow$ PM and (b)
 AFM $\leftrightarrow$ DM phase transitions. \textbf{(a)} The figure in the left (right) panel depicts the variation of 
 $\partial \mathcal{L}/\partial \lambda_1$ ($\partial \mathcal{D}/\partial \lambda_1$) with $\lambda_1$ across the AFM 
 $\leftrightarrow$ PM QPT for different values of $N$, with $\lambda_2=1.5$, and $\gamma=0.8$. (Insets) Corresponding variations of 
 $\ln\left|\lambda_{1}^c(N)-\lambda_{1}^c(\infty)\right|$ (both numerical data and fitted line) as a function of $\ln N$. 
 \textbf{(b)} The figure in the left (right) panel depicts the variation of 
 LN (QD) with $\lambda_2$ across the AFM 
 $\leftrightarrow$ DM QPT for different values of $N$, with $\lambda_1=1.5$. 
 (Insets) Corresponding variations of 
 $\ln\left|\lambda_{2}^c(N)-\lambda_{2}^c(\infty)\right|$ (both numerical data and fitted line) as a function of $\ln N$.  
 All the quantities plotted are dimensionless, 
 except LN which is in ebits and QD, that is in bits.}
 \label{scale}
\end{figure*}

\subsection{Quantum correlations at zero temperature}
\label{stat:zerotemp}

In the limit $\beta\rightarrow\infty$, we now investigate the behavior of LN and QD, as functions of the system 
parameters $\lambda_1$ and $\lambda_2$, in the thermodynamic limit.
For this system, $m_z^e, m_z^o$ and all the non-zero classical correlations can be obtained analytically by diagonalizing $\hat{H}_p$, following the similar prescription for the XY model (see Appendix \ref{ap:corr_fun}) and hence the exact computation of LN and QD is possible, as 
 depicted in the top horizontal pannels of Fig. \ref{lnqdggm}. \textcolor{red}{To keep the notation uncluttered, from now on, we denote LN by $\mathcal{L}$, and QD by $\mathcal{D}$.}  In this paper, all the analysis are carried out 
 for $\gamma = 0.8$ unless specified otherwise. The qualitative feature of the entire 
 investigation remains same for $\gamma \neq 0$. 
Note that the LN has a high value in the DM and PM phases, while the value is low in the AFM phase. On the other hand, the value of 
QD is moderate in the PM region. In the AFM region, the QD has a low value except in the cases where the values of $\lambda_1$ and $\lambda_2$ are 
comparable. Along the line $\lambda_1=-\lambda_2$, QD is vanishingly small. Note that the situation is reversed if one performs 
measurement on the odd qubit while determining QD. In that case, low values of QD are found along the $\lambda_1=\lambda_2$ line. 
Hence, the asymmetry imposed into the model due to the introduction of the alternating field 
is captured from the distribution of QD values over the AFM region in the parameter space of $(\lambda_1,\lambda_2)$, but not by LN. 
However in the case of LN, there exist two zero-entanglement lines in the AFM phase, as depicted in the left pannel of the 
top horizontal row of
Fig. \ref{lnqdggm}, which represent 
fully separable ground states. These lines, which we refer to as the  ``factorization lines'', are discussed in detail in the 
subsequent section. 

One must note here that the introduction of only a local parameter, i.e. the alternating transverse field, in the 
well-known transverse-field XY Hamiltonian \cite{lsm,bmpapers}, gives rise to the DM phase, which is not present in the transverse-field 
XY model. It is interesting to investigate how the QPTs occurring at the AFM $\leftrightarrow$ DM phase boundaries  
can be characterized using entanglement and information-theoretic quantum correlation measures, and whether such characteristic behaviours are 
similar to those observed in the case of the AFM $\leftrightarrow$ PM transition in this model as well as in the usual transverse-field 
XY model \cite{lsm,bmpapers}. In the thermodynamic limit, and for the latter case, 
the QPT is found to be signaled by 
a divergence in the first derivative of entanglement as well as in the information theoretic measures with respect to the system parameters, $\lambda_1$ and 
$\lambda_2$ (see the bottom panel of Fig. \ref{lnqdggm}). We find that, similar to the 
AFM $\leftrightarrow$ PM QPT, other transitions can also be detected by the first derivative of 
appropriate measures of quantum correlations. As an example, in Fig. \ref{lnqdggm} (bottom horizontal pannels), 
we plot $|\partial\mathcal{L}/\partial\lambda_1|$ (left pannel) and $|\partial\mathcal{L}/\partial\lambda_2|$ (right pannel) as functions of
$\lambda_1$, and $\lambda_2$ for $\gamma=0.8$, $\beta\rightarrow\infty$, and $N\rightarrow\infty$. From the figures, 
we can clearly see that both $|\partial\mathcal{L}/\partial\lambda_1|$ (left pannel) and $|\partial\mathcal{L}/\partial\lambda_2|$ 
(right pannel) diverge 
at the AFM $\leftrightarrow$ DM and AFM $\leftrightarrow$ PM boundaries. We plot the absolute values of the first 
derivatives of LN for a better representation of the divergence, as the actual first derivative can tend to both positive as
well as negative infinity, depending on the variation of LN with respect to $\lambda_1$ and $\lambda_2$. Note here that 
there exists two lines, one vertical ($|\partial\mathcal{L}/\partial\lambda_1|$) and the other horizontal 
($|\partial\mathcal{L}/\partial\lambda_2|$) in the variations of the first derivative of LN, as depicted in Fig. \ref{lnqdggm}, over 
which the value of LN remains almost constant. This is indicated  by the low value of the first derivative of LN over those lines.
Note also that there exists several models in which bipartite entanglement cannot detect quantum phase transitions 
\cite{bipartite-failed, amico-rmp, lewenstein-rev}. Such example includes the spin liquid-dimer transition in 1D $J_1-J_2$ 
model \cite{J1-J2}. The results obtained here show that this is not the case for the XY model with uniform and alternating transverse field. 

\subsubsection*{\textbf{Finite-size scaling analysis}}
Advancement of experimental techniques has made the laboratory realization of several quantum many-body systems of finite size, such as
the quantum anisotropic XY model with a transverse alternating magnetic field, possible \cite{lab_xy_ion,solid_xy}, 
which highlights the importance of studying 
the behavior of quantum correlations in the context of QPTs in system of finite number of spins. 
Towards this aim, we present the finite size scaling analysis of the system using the bipartite quantum correlations, and  
determine the scaling exponents. More specifically, we discuss the finite-size scaling of the system at the QPTs corresponding to 
\textbf{(i)} AFM $\leftrightarrow$ PM and \textbf{(ii)} AFM $\leftrightarrow$ DM transitions. 

\noindent\textbf{(i)} \textbf{AFM $\leftrightarrow$ PM transitions:} At $\lambda_2=0$, the model reduces to the widely studied anisotropic 
XY model in a uniform transverse magnetic field of strength $\lambda_1$. As $N\rightarrow \infty$ and $\beta \rightarrow \infty$, 
the model undergoes a QPT, between the quantum PM phase and the 
AFM phase, at $\lambda_1^c=\pm 1$. It is well known that this QPT is signaled by a non-analyticity in the 
first-derivatives of the quantum correlation measures, $\mathcal{Q}$, with respect to the system parameter, $\lambda_1$ \cite{xy-qpt-ent,xy-qpt-disc}.  
With the introduction of the transverse alternating field $\lambda_2$, the QPT point changes according to the line $\lambda_1^2=\lambda_2^2+1$, 
which denotes the phase-boundary between the AFM and the quantum PM phase in the present model. As shown in Fig. \ref{lnqdggm}, 
the AFM $\leftrightarrow$ PM transition is also signalled by a non-analyticity in the first-derivative of LN, or QD, with respect to 
$\lambda_1 (\lambda_2)$, when $\lambda_2 (\lambda_1)$ is kept fixed as $N\rightarrow \infty$. In the case of a system of finite size, the QPT is signalled by a maximum or a minimum in 
the variation of the first-derivative of LN and QD with respect to $\lambda_1 (\lambda_2)$, for fixed values of $\lambda_2 (\lambda_1)$ (see Fig. \ref{scale}). The position of the maximum
or minimum denotes the position of the critical point on the axes of the respective system parameter. The maximum or minimum sharpens with 
increasing system size, and the position of the QPT approaches the QPT point as $N \rightarrow \infty$, denoted by $\lambda^c_{1(2)}(\infty)$, as 
\begin{eqnarray}
 \lambda_{1(2)}^c(N)=\lambda_{1(2)}^c(\infty)+\alpha_{1(2)}N^{-\nu_{1(2)}}.
 \label{scale_eq}
\end{eqnarray}
Here, $\alpha_{1(2)}$ are dimensionless constants, and $\nu_{1(2)}$ are the scaling exponents. 

Fig. \ref{scale}(a) depicts the variation of 
derivative of LN and QD, as functions of $\lambda_1$, as $\beta\rightarrow\infty$, for fixed value of $\lambda_2$ i.e. $\lambda_2 = 1.5$ with 
$\gamma=0.8$. The approach of the QPT points, $\lambda_1^c(N)$, 
at finite $N$, towards the QPT point in the thermodynamic limit, $\lambda_1^c(\infty)$, are depicted in the insets. Fitting the numerical data
with Eq. (\ref{scale_eq}), one can estimate the values of $\alpha_1$ and $\nu_1$. Table \ref{expo_afmpm}(a) contains the values of $\alpha_{1,2}$
and $\nu_{1,2}$, in the case of both LN and QD, when the value of $\lambda_2$ $(\lambda_1)$ is kept fixed at $\lambda_2=0$ and $1.5$ ($\lambda_1=1.5$). Note that 
the values of $\alpha_{1,2}$ and $\nu_{1,2}$ change
with $\gamma$ although the qualitative feature remains invariant. 

\begin{table}

\begin{tabular}{|c|}
   \hline 
   Tuning parameter: $\lambda_1$\\
   \hline 
   \begin{tabular}{c|c|c}
      $\lambda_2$ & LN & QD\\
      \hline
      $0.0$ & \begin{tabular}{c}
                 $\nu_1=1.645 \pm 0.013$\\
                 $\ln\alpha_1=2.842 \pm 0.070$ \\
              \end{tabular}
            & \begin{tabular}{c}
                 $\nu_1=1.292 \pm 0.093$\\
                 $\ln\alpha_1=1.851 \pm 0.631$ \\
              \end{tabular}\\
      \hline
      $1.5$ & \begin{tabular}{c}
                 $\nu_1=2.278  \pm  0.053$\\
                 $\ln\alpha_1=3.828 \pm  0.230$ \\
              \end{tabular}
            & \begin{tabular}{c}
                 $\nu_1=1.489 \pm  0.027$\\
                 $\ln\alpha_1=1.507 \pm 0.175$ \\
              \end{tabular}\\
   \end{tabular}\\
   \hline
\end{tabular}

\vspace{0.5cm}

\begin{tabular}{|c|}
   \hline 
   Tuning parameter: $\lambda_2$\\
   \hline 
   \begin{tabular}{c|c|c}
      $\lambda_1$ & LN & QD\\
      \hline
      $1.5$ & \begin{tabular}{c}
                 $\nu_2=1.941 \pm 0.042$\\
                 $\ln\alpha_2=3.614 \pm 0.204$ \\
              \end{tabular}
            & \begin{tabular}{c}
                 $\nu_2=1.507 \pm  0.008$\\
                 $\ln\alpha_2=2.380 \pm  0.052$ \\
              \end{tabular}\\
   \end{tabular}\\
   \hline
\end{tabular}
\caption{Finite-size scaling exponents and fitting parameters for the QPT corresponding to AFM $\leftrightarrow$ PM transition.
For all the computations, $\gamma=0.8$.}
\label{expo_afmpm}
\end{table}
%
%
%

\noindent\textbf{(ii) AFM $\leftrightarrow$ DM transition:} Similar to the case of AFM $\leftrightarrow$ PM transition, in the thermodynamic 
limit, the AFM $\leftrightarrow$ DM transition is signaled by a non-analyticity in the first-derivative of LN, or QD,
with respect to either of $\lambda_1$ and $\lambda_2$. 
It is interesting to investigate how the position of the QPT point, as determined by the position of the sharp peak in 
the variation of the derivatives of 
LN and QD with respect to either $\lambda_1$, or $\lambda_2$, changes with a variation in the system size. In order to do so, 
one may try to determine the canonical equilibrium state at zero 
temperature by using the same methodology as in the case of the AFM $\leftrightarrow$ PM transition.  
However, due to the approximations involved in determining the zero-temperature state, in the present case,  
LN and QD, as functions of either of $\lambda_1$ and $\lambda_2$, exhibit finite jumps in values at the QPT point,
thereby forbidding a finite-size analysis in a similar fashion as in the previous case
(for a discussion on the behaviour of the finite jumps, and a figure, see Appendix \ref{afmdm}). 
Therefore, we employ the exact diagonalization technique in the present case, and determine the non-degenerate ground state of the Hamiltonian 
given in Eq. (\ref{ham_spin}) by using Lanczos algorithm \cite{lanczos}.  
The reduced density matrix correspponding to a nearest-neighbour even-odd spin pair, labeled by ``$eo$'', can be determined
by tracing out all the other spin variables from the ground state. Using the reduced density matrix,
the nearest-neighbour LN and QD can be computed. Here, for the purpose of discussions, the first-derivatives of 
LN and QD, with respect to $\lambda_2$, by keeping $\lambda_1$ fixed at $1.5$, 
are plotted in Fig. \ref{scale}(b). We find that in the case of the AFM $\leftrightarrow$ DM transition also, the position of the QPT 
at a finite $N$ approaches the actual QPT point at $N\rightarrow\infty$ according to an equation similar to Eq. (\ref{scale_eq}), 
where the constants are denoted by $\alpha_{1,2}$ and $\nu_{1,2}$. For example, for $\lambda_1=1.5$, 
the corresponding values of these fitting parameters are $\nu_2= 2.525 \pm 0.084$, $\ln{\alpha}_2=2.077 \pm 0.220$ (for LN), and $\nu_2= 1.153 \pm 0.036$, $\ln{\alpha}_2=-0.568 \pm 0.092$ (for QD).


\subsection{Factorization line: Separable ground state}
\label{stat:factor}

We now discuss the occurrence of the separable ground state in the AFM phase of the model which can observed by considering the variation of bipartite as well as multipartite entanglement as functions of $\lambda_1$ and $\lambda_2$ (Fig. \ref{lnqdggm}). The symmetry of the Hamiltonian (Eq. (\ref{ham_spin}))
under PBC motivates one to look for a separable eigenstate of the form
\begin{eqnarray}
 |\psi\rangle=\prod_{i=0}^{\frac{N}{2}-1}|\psi_{2i+1}^o\rangle\otimes|\psi_{2i+2}^e\rangle,
 \label{eigs_fact}
\end{eqnarray}
having a N\'{e}el type order, where $|\psi_{\alpha}^{o(e)}\rangle$ are the states of the spins on the odd (even) site $\alpha$. 
The Hamiltonian can be written as $H=\sum_{i=0}^{(N/2)-1}(H^{oe}_{2i+1,2i+2}+H^{eo}_{2i+2,2i+3})$, 
where $H^{oe}$ is the two-site Hamiltonian given by
\begin{eqnarray}
H^{eo}&=&J\left\{\frac{1+\gamma}{4}\sigma_e^x\sigma_{o}^{x}
+\frac{1-\gamma}{4}\sigma_e^y\sigma_{o}^{y}\right\}+\frac{h_+}{2}\sigma_e^z+\frac{h_-}{2}\sigma_o^z
,\nonumber\\
\label{ham_oe}
\end{eqnarray}
with $h_\pm=h_1\pm h_2$, defined on an even-odd pair of sites, and $H^{oe}$ can be obtained from $H^{eo}$ 
straightforwardly by interchanging the site indices. 
Using Eq. (\ref{eigs_fact}), the lowest separable eigenenergy 
can be obtained as 
\begin{eqnarray}
 E_{\mbox{\scriptsize min\normalsize}}^{\mbox{\scriptsize sep\normalsize}}
 &=&\underset{|\psi^e\rangle,|\psi^o\rangle}{\min}\langle\psi|H|\psi\rangle\nonumber \\
 &=&\sum_{i=0}^{\frac{N}{2}-1}\underset{|\psi^e\rangle,|\psi^o\rangle}{\min}\langle\psi^e|\langle\psi^o|H^{eo}_{2i+1,2i+2}|\psi^e\rangle|\psi^o\rangle\nonumber\\
 &&+\sum_{i=0}^{\frac{N}{2}-1}\underset{|\psi^o\rangle,|\psi^e\rangle}{\min}\langle\psi^o|\langle\psi^e|H^{oe}_{2i+2,2i+3}|\psi^o\rangle|\psi^e\rangle\nonumber\\
 &=& N\underset{|\psi^e\rangle,|\psi^o\rangle}{\min}\langle\psi^e|\langle\psi^o|H^{eo}|\psi^e\rangle|\psi^o\rangle,\nonumber 
\end{eqnarray}
where we have used the fact that $H^{eo}$ and $H^{oe}$ are energetically equivalent. This leads to a minimum separable energy per site, 
$\epsilon$, given by $\epsilon=\underset{|\psi^e\rangle,|\psi^o\rangle}{\min}\langle\psi^e|\langle\psi^o|H^{eo}|\psi^e\rangle|\psi^o\rangle$.
Without any loss of generality, one can choose the states $|\psi^{e(o)}\rangle$ to be 
\begin{eqnarray}
 |\psi^{e(o)}\rangle=\cos\frac{\theta_{e(o)}}{2}|0\rangle+\exp{i\phi_{e(o)}}\sin\frac{\theta_{e(o)}}{2}|1\rangle,
\end{eqnarray}
where $\theta_{e(o)}$ and $\phi_{e(o)}$ are real parameters such that $0\leq\theta_{e(o)}\leq\pi$ and $0\leq\phi_{e(o)}\leq 2\pi$. The 
Two-spin reduced density matrix $\rho_{eo}$, corresponding to the odd-even pair of spins, is then given by 
$\rho_{eo}=\rho_e\otimes\rho_o$, where $\rho_{e(o)}=|\psi^{e(o)}\rangle\langle\psi^{e(o)}|$.  Since the Hamiltonian in Eq. (\ref{ham_spin}) is a 
real one, we expect $\rho_{eo}=\rho_{eo}^*$, leading to 
\begin{eqnarray}
 \epsilon&=&\underset{\theta_e,\theta_o}{\min}\frac{1}{4}
 \Big\{J(1+\gamma)\sin\theta_e\sin\theta_o+h_+\cos\theta_e+h_-\cos\theta_o\Big\},\nonumber\\
 \label{epsi}
\end{eqnarray}
where the optimization over the states $|\psi^{o(e)}\rangle$ is reduced to an optimization over the real parameter space of $\theta_o$ and 
$\theta_e$. The minimum is achieved for 
\begin{eqnarray}
\theta_e&=&\tan^{-1}\left\{\pm\frac{1}{h_+}\sqrt{\frac{J^4(1+\gamma)^4-h_+^2h_-^2}{J^2(1+\gamma)^2+h_+^2}}\right\} \nonumber \\
 \theta_o&=&\tan^{-1}\left\{\pm\frac{1}{h_+}\sqrt{\frac{J^4(1+\gamma)^4-h_+^2h_-^2}{J^2(1+\gamma)^2+h_-^2}}\right\}.
 \label{thetamin}
\end{eqnarray}

However, the state $|\psi\rangle$ (Eq. (\ref{eigs_fact})) would be the ground state of the Hamiltonian if 
$\epsilon=\epsilon_0$, the ground state energy of the two-spin 
Hamiltonian $H^{eo}$ \cite{factor_old,factor_new}. We find that the ground state of $H^{eo}$ is nondegenerate, with a ground state energy given 
by $\epsilon_0=\frac{1}{2}\sqrt{J^2+4h_2^2}$. Determination of $\epsilon$ using the values of 
$\theta_{o(e)}$, and equating to $\epsilon_0$ leads to the following condition 
\begin{eqnarray}
 h_1^2=h_2^2+J^2(1-\gamma^2),
 \label{eq_factor}
\end{eqnarray}
equivalently $\lambda_1^2=\lambda_2^2+(1-\gamma^2)$, which represents a line on the $(\lambda_1,\lambda_2)$ plane for fixed values of $\gamma$. The ground state of the Hamiltonian, at 
every point on this line on the $(\lambda_1,\lambda_2)$ plane, is separable, represented by a line of vanishing entanglement (Fig. \ref{lnqdggm}).  We call this line as {\it factorization line}.  \textcolor{red}{At any point on this line, the minimum eigenvalue of $\hat{H}_p$, given by  $-\omega^4_+(p)$, becomes independent of  $\phi_p$, as demonstrated  for ($\lambda_1=0.6, \lambda_2=0$) with $\gamma=0.8$ in Fig. \ref{spectrum}(c). This feature is in contrast to the $\phi_p$ dependence of $-\omega^4_+(p)$ at the QPT points (see Figs. \ref{spectrum}(a) and (b)).}

\begin{figure}
 \includegraphics[scale=0.3175]{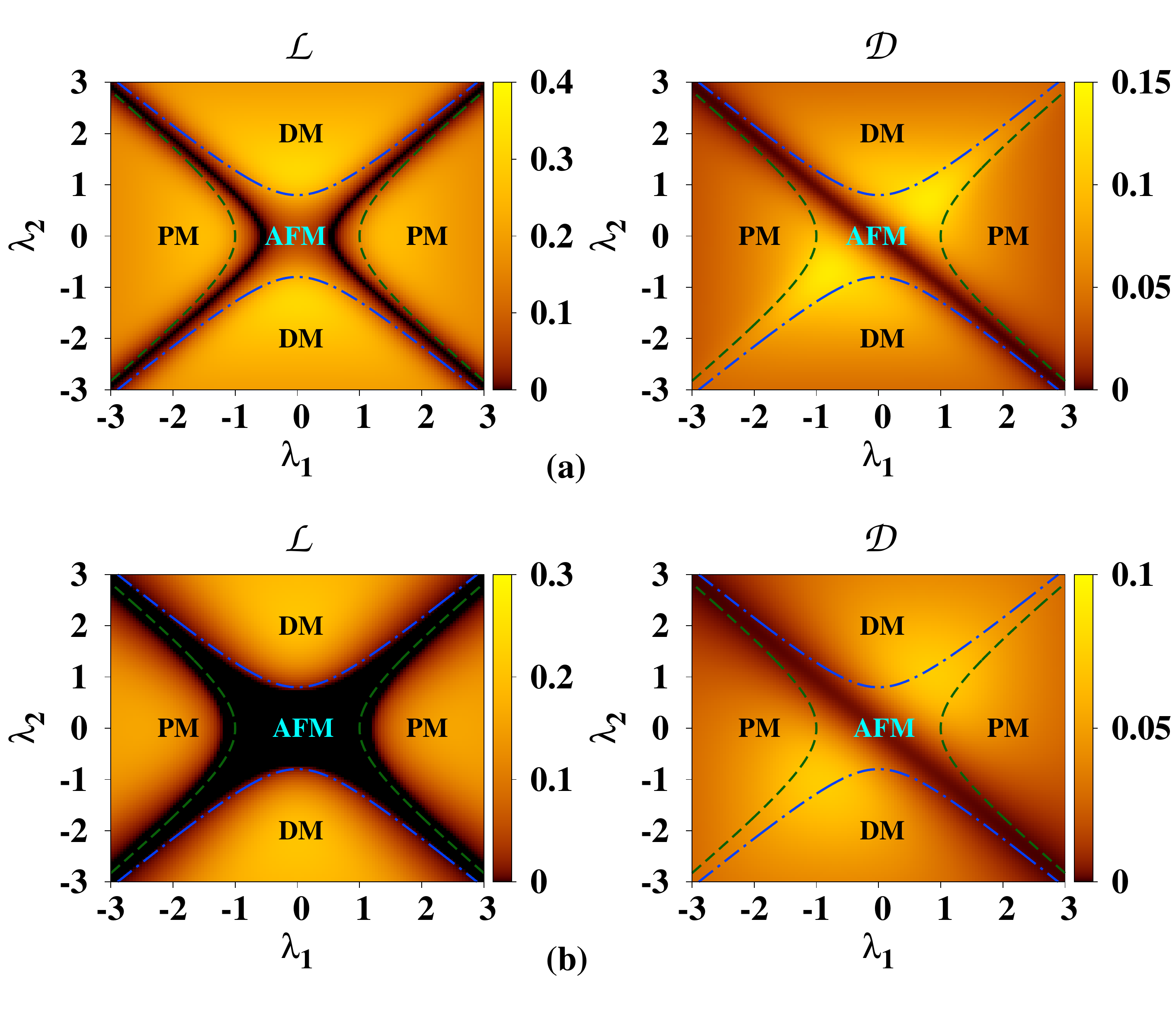}
 \caption{(Color online) Variation of LN and QD as functions of the transverse magnetic field $h_1$ and the alternating field $h_2$
 in the thermodynamic limit at (a) $\beta J=5$ and (b) $\beta J=2$, and $\gamma=0.8$. 
 The zero-temperature phase boundaries, $\lambda_1^2=\lambda_2^2+1$ (PM $\leftrightarrow$ AFM) and 
 $\lambda_2^2=\lambda_1^2+\gamma^2$ (AFM $\leftrightarrow$ DM), are also plotted for comparison, 
 represented by the dashed and dot-dashed lines, respectively.   All the quantities plotted are dimensionless, except LN and QD, which are in ebits and bits respectively.}
 \label{temp_phase}
\end{figure}

\subsection{Effect of temperature on quantum correlations}
\label{stat:temp}

Quantum correlations are known to be fragile quantities, and are expected to decay with increasing thermal noise in the system. Moreover, absolute zero temperature is hard to be achieved in a real experiment. It is therefore 
interesting to investigate the effect of thermal fluctuations on the bipartite  quantum correlations corresponding to the 
Hamiltonian (Eq. (\ref{ham_spin})). The patterns of LN and QD as functions of $\lambda_1$ and $\lambda_2$, for
$\beta J=5$ (Fig. \ref{temp_phase}(a)) and $\beta J=2$ (Fig. \ref{temp_phase}(b)) are plotted in Fig. \ref{temp_phase}. In the case of LN, 
we observe that starting from the factorization line at $\beta\rightarrow\infty$, a zero-entanglement region grows with increasing 
temperature, and spans the entire AFM phase at sufficiently high temperature. Note here that the zero-entanglement region at $\beta J>0$ 
can also be found in the PM phase, while it is absent in the DM phase even at high temperature. A few interesting features emerge 
from these results.

\begin{figure}
 \includegraphics[scale=0.355]{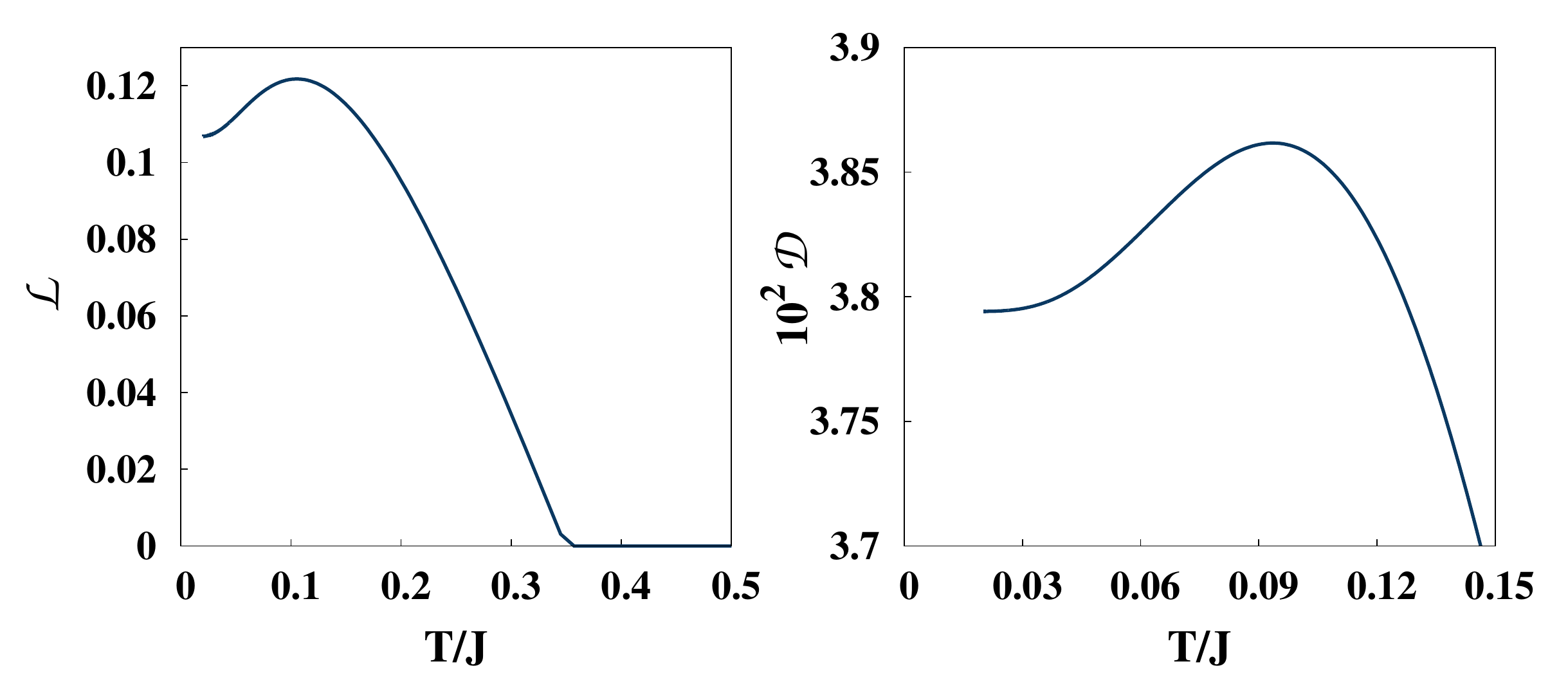}
 \caption{(Color online) Non-monotonic variations of LN (left panel) and QD (right panel) with temperature. We choose  $\lambda_1=-0.9,\lambda_2=0.25$ for LN, and $\lambda_1=-0.4,\lambda_2=0.7$ for QD. Here, $\gamma=0.8$. LN and QD are measured respectively in ebits and bits. T/J is dimensionless, 
}
 \label{nonmono}
\end{figure}

\begin{enumerate}
\item[\textbf{(i)}] The rate of spreading of the vanishing entanglement region with increasing temperature is found to be much slower towards 
the PM phase compared to that inside the AFM phase. This can be easily perceived from the fact that with the increase of temperature from 
$\beta J=5$ to $\beta J=2$, the entanglement vanishes in the entire AFM phase, but covers only a small region in the PM phase. 
It implies that bipartite entanglement is more fragile in the AFM region compared to the other phases. 

\item[\textbf{(ii)}] Remarkably, bipartite entanglement in the DM phase is the most robust against increasing thermal noise 
among the three phases.

\item[\textbf{(iii)}] The effect of thermal noise on QD is less drastic compared to that in the case of LN, as observed from Fig. 
\ref{temp_phase}. With increasing temperature, the minimum value of QD along the line $\lambda_1=-\lambda_2$ increases. However, the qualitative 
distribution of QD over the $(\lambda_1,\lambda_2)$ plane remains unchanged. 
\end{enumerate}

\noindent\textbf{Remark 1.} We choose $\beta J=2.0$ and treat as high temperature since bipartite entanglement of the AFM phase has been destroyed at this temperature. However, if one increases temperature beyond $\beta J=2$, 
LN in the entire region of $(\lambda_1,\lambda_2)$ plane becomes zero. 

\noindent\textbf{Remark 2.} For the purpose of demonstration, we have kept the anisotropy parameter constant to a fixed value $\gamma=0.8$. 
One must remember that the definition of ``high'' $\beta J$ depends, along with the other system parameters, on the anisotropy parameter also. 
However, the qualitative features, such as the robustness of bipartite entanglement 
in the DM phase compared to other phases, or the fragility of LN in the AFM phase remain unchanged 
with a change in the value of the anisotropy parameter.

\begin{figure*}
 \includegraphics[width=\textwidth]{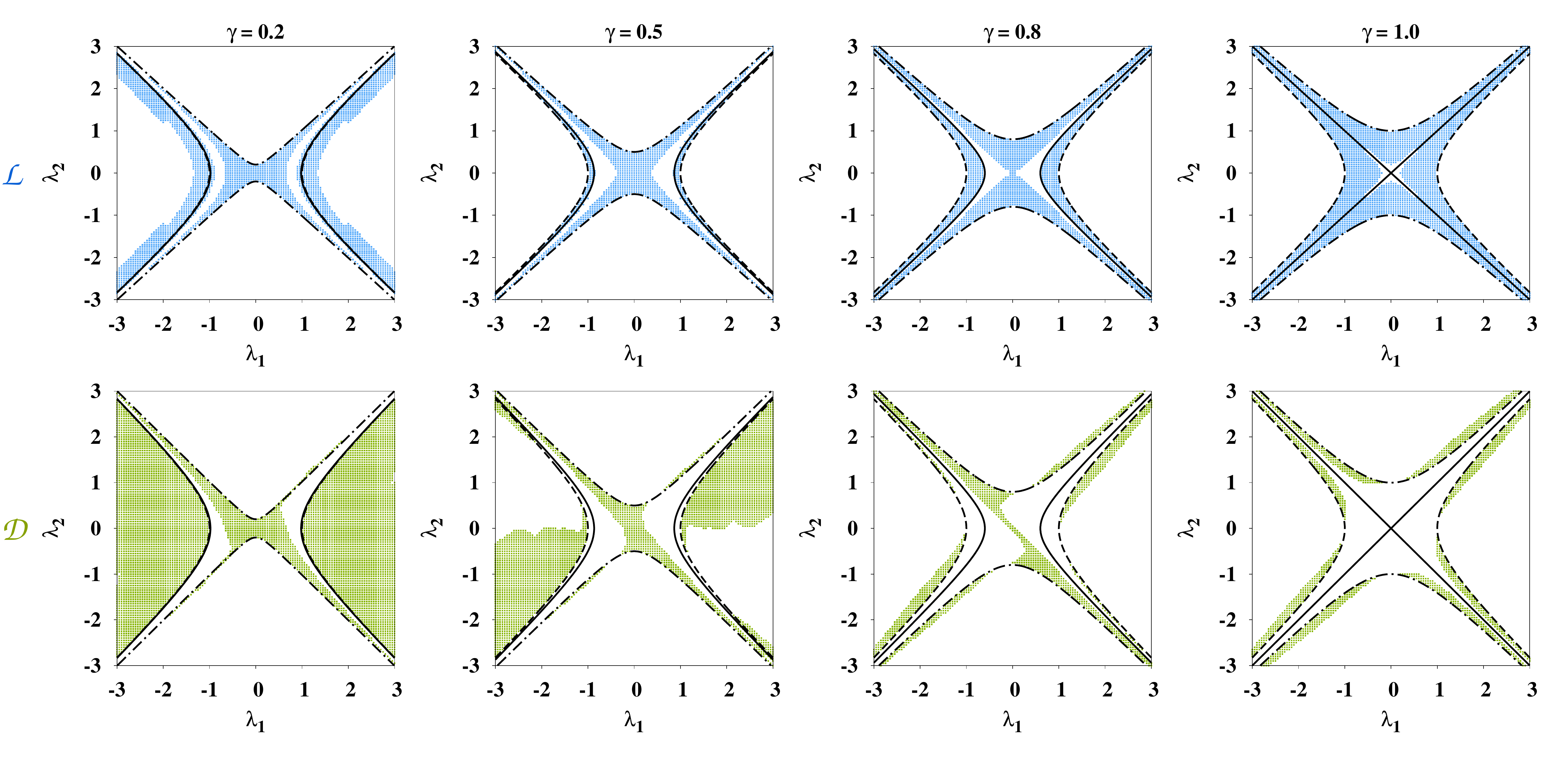}
 \caption{(Color online) Map of the regions over the $(\lambda_1,\lambda_2)$ plane where non-monotonic behavior of LN (figures in top horizontal panels) and QD (figures in bottom horizontal panels), with variation of temperature 
 (marked by the shaded regions). The zero-temperature phase boundaries, $\lambda_1^2=\lambda_2^2+1$ (PM $\leftrightarrow$ AFM), 
 $\lambda_2^2=\lambda_1^2+\gamma^2$ (AFM $\leftrightarrow$ DM), and the factorization line  ($\lambda_1^2=\lambda_2^2+(1-\gamma^2)$) are also plotted for comparison, 
$\lambda_1$ and $\lambda_2$ are dimensionless while LN and QD are in  ebits and bits respectively.
}
 \label{mono}
\end{figure*}

\subsubsection*{\textbf{Monotonicity vs. non-monotonicity}}

Up to now we have discussed the variation in the pattern of entanglement and QD with the increase of temperature. We now report 
 the existence of non-monotonic variation of LN and QD as functions of temperature in this model. Such non-monotonicity is 
 known for other quantum many-body Hamiltonians, including the transverse-field XY model \cite{other-nonmono,other-qpt-ent,other-qpt-ent-2}.
 Since the model under consideration possess different phase diagram than the XY model, 
 non-monotonicity of quantum correlations, specially  entanglement with temperature may reveal some new feature. 
 We will show that this is indeed the case. 
  For fixed choices of $(\lambda_1, \lambda_2)$, typical variation profiles exhibiting 
non-monotonicity of LN and QD with temperature, as shown in Fig \ref{nonmono}. 
The importance of non-monotonic behavior of bipartite quantum correlation lies in the fact that even at high temperature, which is much easier to attain in the laboratory, a higher value of  
 quantum correlations is obtained compared to the state with lower temperature.
   This has potential applicability in the realization of those quantum 
protocols in the laboratory, which use quantum correlations as resources.

 It is therefore necessary to map the occurrence of non-monotonic
variations of bipartite quantum correlations over the phase plane of the model, so that the useful regions at finite temperature can be 
recognized. Let us consider a set of values in the space of the system parameters, denoted by $\{\lambda_1,\lambda_2,\gamma\}_{\mathcal{Q}}$, 
which results in a non-monotonic variation of the bipartite quantum correlation measure, $\mathcal{Q}$, with the variation of temperature. 
We call such a set as the ``non-monotonicity generator'' (NG). 
Fig. \ref{mono} exhibits the NGs for different values of $\gamma$, specially $\gamma=0.2$, $0.5$, $0.8$, and $1.0$, 
 on the $(\lambda_1,\lambda_2)$-plane, when LN and QD are considered to be the bipartite quantum correlation measures. 
We observe that in the case of LN, for low values of $\gamma$, the NGs are confined to the AFM phase, and narrow 
regions inside the PM phase, in the vicinity of the AFM $\leftrightarrow$ PM QPT line. At $\gamma=0.2$, the factorization lines, denoted by the 
solid line on the $(\lambda_1,\lambda_2)$ plane, almost coincides with the AFM $\leftrightarrow$ PM QPT line, which is represented by the dashed lines. 
With increasing value of $\gamma$, the factorization lines get separated from the AFM $\leftrightarrow$ PM transition lines, and the NGs span 
the region confined by these lines, as can be seen in the case of $\gamma=0.5$ and $\gamma=0.8$. At $\gamma=1.0$, which represents the 
Ising model in transverse-uniform and transverse-alternating field, the factorization lines meet each other, and almost entire AFM phase is filled by the NGs. 
Remarkably, the DM phase remains completely free from NGs for all values of $\gamma$.

\begin{figure}
 \includegraphics[scale=0.35]{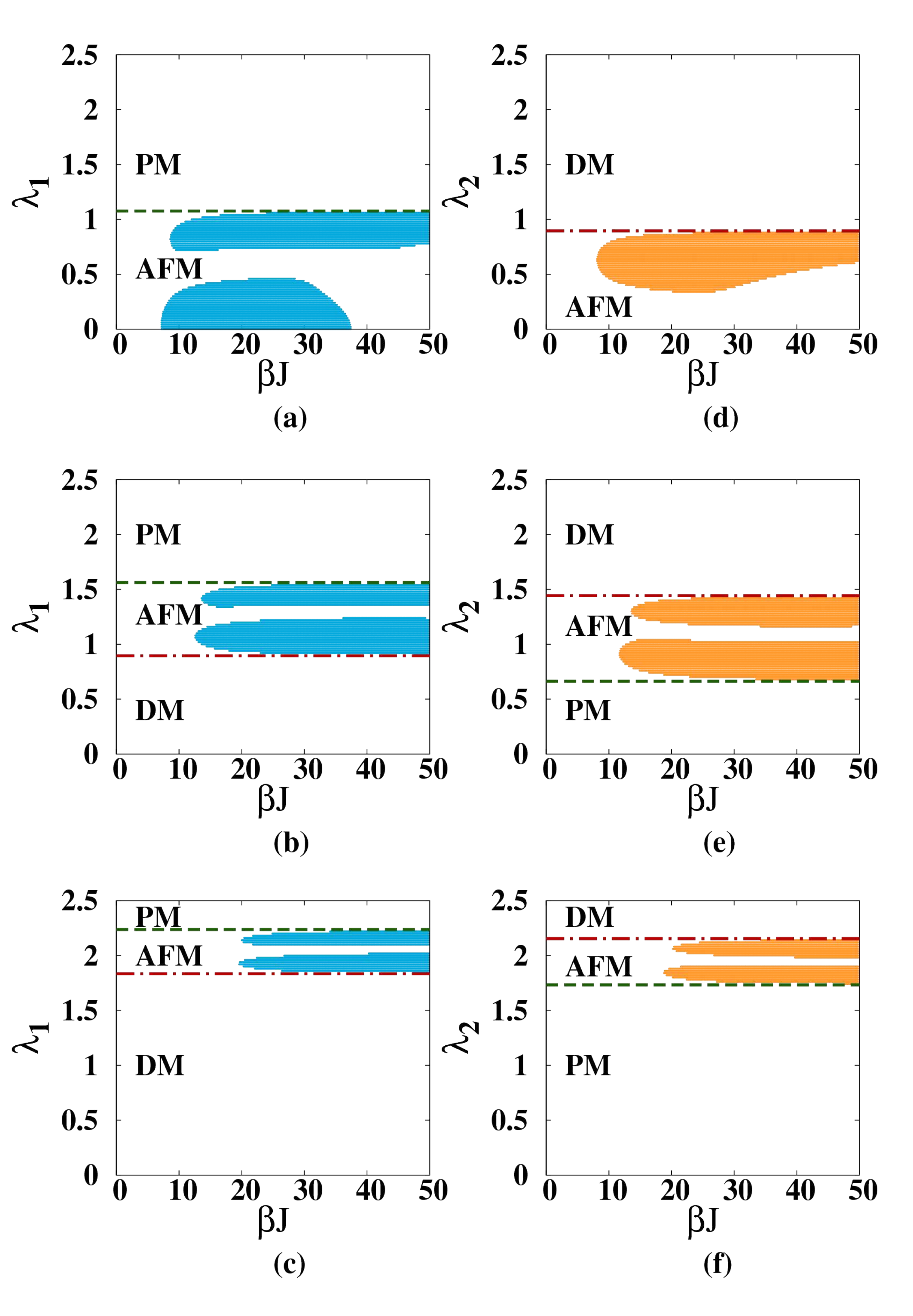}
 \caption{(Color online) Map of the regions on the $(\lambda_{1(2)},\beta J)$ plane, where LN decreases with an increasing $\beta J$, 
 with the value of $\lambda_{2(1)}$ being fixed at \textbf{(a)}, \textbf{(d)} $\lambda_{2(1)}=0.4$, \textbf{(b)}, \textbf{(e)} $\lambda_{2(1)}=1.2$, and 
 \textbf{(c)}, \textbf{(f)} $\lambda_{2(1)}=2.0$.
All the other lines are same as Fig. \ref{mono}.
Here, $\lambda_1$ and $\lambda_2$ are dimensionless while $\beta J$ has the dimension of energy with $k_{B} = 1.$
}
 \label{nonmono_beta}
\end{figure}


The behaviors of QD and LN, with respect to non-monotonicity, are somewhat 
complementary to each other for low and high values of the anisotropy parameter. At $\gamma=0.2$, NGs for QD span the PM phase, which is in contrast to 
the case of LN, where NGs can be found in the PM phase only in the vicinity of the AFM $\leftrightarrow$ PM phase boundary. On the other hand, 
for $\gamma=1.0$, in the case of LN, NGs fill almost entire AFM phase while being absent in the PM and the DM phase while in the case of QD, 
non-monotonicity occurs in a very small region of the AFM and DM phase.
In Fig. \ref{nonmono_beta}, we map, on the $(\lambda_{1(2)},\beta J)$-plane, the regions where LN increases  
with decreasing the value of $\beta$ which confirms the findings in Fig.  \ref{mono}.   To generate the figures corresponding to the $(\lambda_{1(2)},\beta J)$-plane, we have kept the value of 
$\lambda_{2(1)}$ fixed.

\noindent\textcolor{red}{\textbf{Note.} In a system of finite number of spin-$1/2$ particles, use of the open boundary condition (OBC) instead of the PBC changes the phase boundaries only slightly, and the AFM region on the $(\lambda_1,\lambda_2)$ plane shrinks. With an increase in the system size, the difference between the phase portraits corresponding to the PBC and the OBC reduces. Note here that each of the pairs of nearest-neighbour spins in the quantum spin model described by Eq. (\ref{ham_spin}) consist of an even, and an odd spin. In the case of the PBC, there is a special type of translational symmetry in the model, such that $\rho_{i,i+1}=\rho_{i+2,i+1}$, where $i$ is, say, an odd site. Hence, LN is same for all the nearest-neighbour spin pairs, while due to this property, quantum discord is same only when measurement is performed on the same type of spin (even or odd) in all nearest-neighbour spin pairs. This implies that under PBC, investigation of the bipartite quantum correlations belonging to any one of the nerest-neighbour spin pairs suffice. On the other hand,  for complete  characterization of the static and dynamical behaviour of nearest-neighbour  bipartite quantum correlations in a system of $N$ spins under OBC, computation of bipartite quantum correlation measures corresponding to $N/2$ $((N-1)/2)$ nearest-neighbour pairs, depending on whether $N$ is even (odd), is necessary. However, the broad qualitative features of the factorization line and the phase boundaries, as reported in this paper, remain unaltered even under OBC for finite-sized systems.}

\section{Dynamics of quantum correlations}
\label{dynamic}

So far, we have considered the static characteristics of quantum correlations in different phases of the 1d anisotropic XY model in uniform and alternating transverse field. 
In this section, we aim to study the behaviour of quantum correlations and their  statistical mechanical properties  under time evolution. 
In order to compute nearest neighbour LN and QD of TES, the two-spin reduced density matrix has 
to be determined, which, in turn, requires the evaluation of single-site magnetizations and two-site spin correlation functions. 
This can be done by utilizing the fact that the evolutions of the subspaces in the momentum space (see Sec. \ref{model} and  Appendix 
\ref{ap:diagonalization}) are independent of each other. This leads to $\hat{\rho}^p(t)=e^{-i\hat{H}_pt}\hat{\rho}^p(0)e^{i\hat{H}_pt}$, where 
$\hat{H}_p$ is the Hamiltonian in the $p^{th}$ momentum subspace at $t>0$, and 
$\hat{\rho}^p(0)=\hat{\rho}^p_{eq}(0)$.  The time-evolved single-site magnetizations and two-site spin correlation functions are given by
\begin{eqnarray}
 m^z_{o(e)}(t)&=&\frac{2}{N}\sum_{p=1}^{N/4}\mbox{Tr}[\hat{m}_p^{z,o(e)}\hat{\rho}^p(t)]/\mbox{Tr}[\hat{\rho}^p(t)],\nonumber\\
 c^{\alpha\beta}(t)&=&\frac{2}{N}\sum_{p=1}^{N/4}\mbox{Tr}[\hat{c}_p^{\alpha\beta}\hat{\rho}^p(t)]/\mbox{Tr}[\hat{\rho}^p(t)].
\label{mtcpt}
\end{eqnarray}
Note here that Eq. (\ref{mtcpt}) addresses systems of finite size, $N$. In the thermodynamic limit, the relevant quantities are obtained by 
replacing the sum with an proper integral, as discussed in Sec. \ref{static_qc}. Note also that unlike the CES, 
$c^{xy}(t)$ and $c^{yx}(t)$ corresponding to 
TES do not vanish, which leads to a contribution in the $c^{zz}$, given by 
\begin{eqnarray}
 c^{zz}(t)=m^z_o(t)m^z_e(t)-c^{xx}(t)c^{yy}(t)+c^{xy}(t)c^{yx}(t).
\end{eqnarray}

\subsection{Ergodicity and ergodicity score}

\begin{figure}
 \includegraphics[scale=0.35]{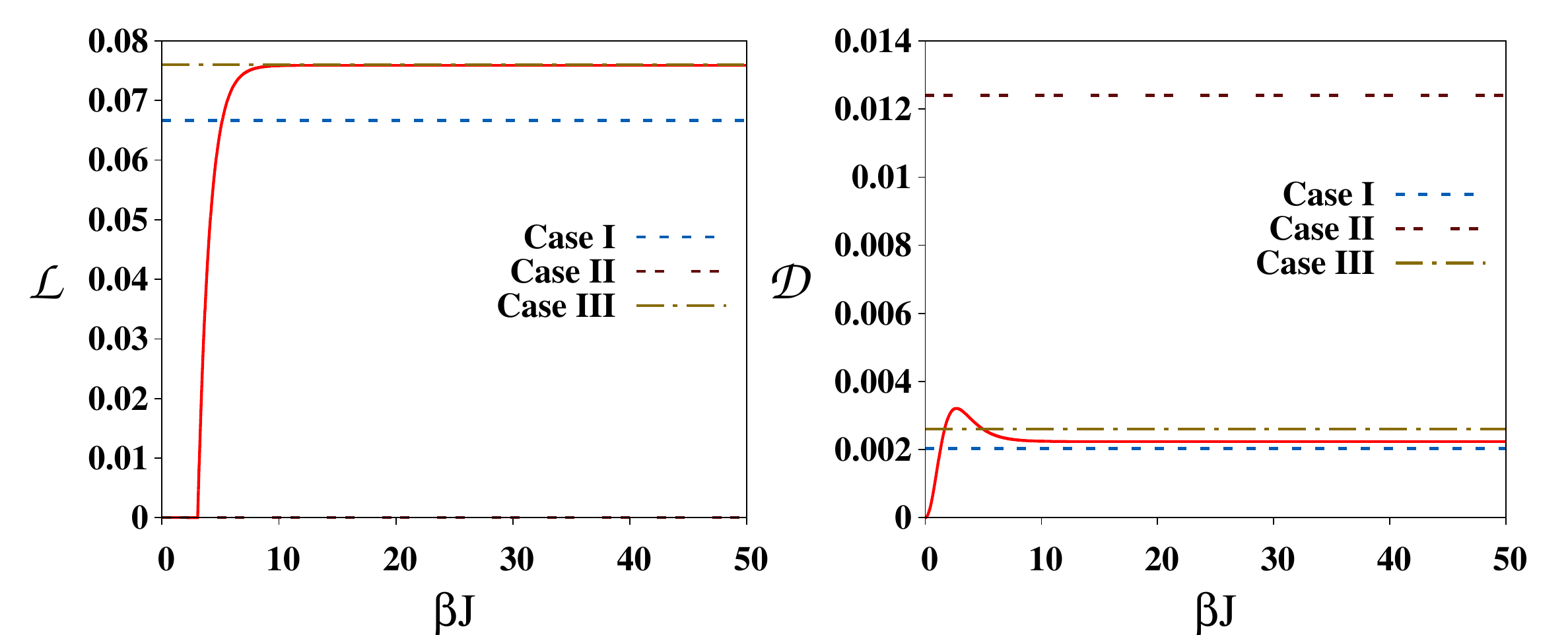}
 \caption{(Color online) Ergodicity of LN and QD for three specific cases. \textbf{Case I}: $(h_1=0.0,h_2=0.15)$. 
 Here, both LN and QD are ergodic. \textbf{Case II}: $(\lambda_1=1.0,\lambda_2=2.0)$. Here, LN is ergodic, but QD is nonergodic.
 \textbf{Case III}: $(\lambda_1=0.15,\lambda_2=0.0)$. Here, QD is clearly ergodic, while the status of LN is 
 inconclusive, and depends on the numerical accuracy.
 For all the cases, initial temperature at $t=0$ is taken to be $\beta J=100$ 
 at $\gamma=0.8$. 
 LN and QD are in ebits and bits respectively while $\beta J$ is  dimensionless.
 }
 \label{ergoinst}
\end{figure}

Let us now discuss the statistical mechanical properties, specifically the ergodicity of bipartite quantum correlations, in the case of the Hamiltonian given in 
Eq. (\ref{ham_spin}). We start with a brief description and quantification of ergodicity of a generic quantum correlation measure, $\mathcal{Q}$.
A physical quantity is said to be ergodic if the time average of the quantity is the same as its ensemble average. In the present scenario, 
the bipartite quantum correlation, $\mathcal{Q}$, is said to be ergodic if there exists a temperature, $T$, at which the ``large time'' 
time-averaged value of $\mathcal{Q}$ in the TES, given by $\mathcal{Q}_{\infty}(T,\lambda_1,\lambda_2)$,  coincides with 
$\mathcal{Q}_{eq}(T^\prime,\lambda_1^{\infty},\lambda_2^{\infty})$, the value of $\mathcal{Q}$
in the CES at temperature $T^\prime$ at $t\rightarrow\infty$. Here, $\lambda_{1(2)}(t\rightarrow\infty)=\lambda_{1(2)}^\infty$. We shall shortly 
discuss what we mean by ``large'' time. 
Using the above definitions, one can define an ``ergodicity score'' as \cite{dyn-ergo-xy-group,dyn-ergo-xy-dual}
\begin{eqnarray}
 \eta_{\mathcal{S}}^{\mathcal{Q}}=\max\left[0,\mathcal{Q}_{\infty}(T,\lambda_1,\lambda_2)
 -\underset{T^\prime}{\max}\mathcal{Q}_{eq}(T^\prime, \lambda_1^{\infty},\lambda_2^{\infty})\right],\nonumber \\
 \label{ergo_score}
\end{eqnarray}
where $\mathcal{S}$ is the set of all system parameters, $\{\lambda_1,\lambda_2,\gamma\}$, and the maximization inside the parenthesis is over 
the physically relevant range of $T^\prime$, which is up to an order of magnitude of $T$. Note that the value of the ergodicity score depends on all the  
relevant system parameters, viz. $\lambda_1$, $\lambda_2$ and $\gamma$, which is indicated by the subscript $\mathcal{S}$. As evident from 
the definition, a non-zero value of $\eta_{\mathcal{S}}^{\mathcal{Q}}$ implies the non-ergodicity of $\mathcal{Q}$, 
while the vanishing $\eta_{\mathcal{S}}^{\mathcal{Q}}$ indicates that the quantity is ergodic.

\begin{figure}
 \includegraphics[scale=0.3135]{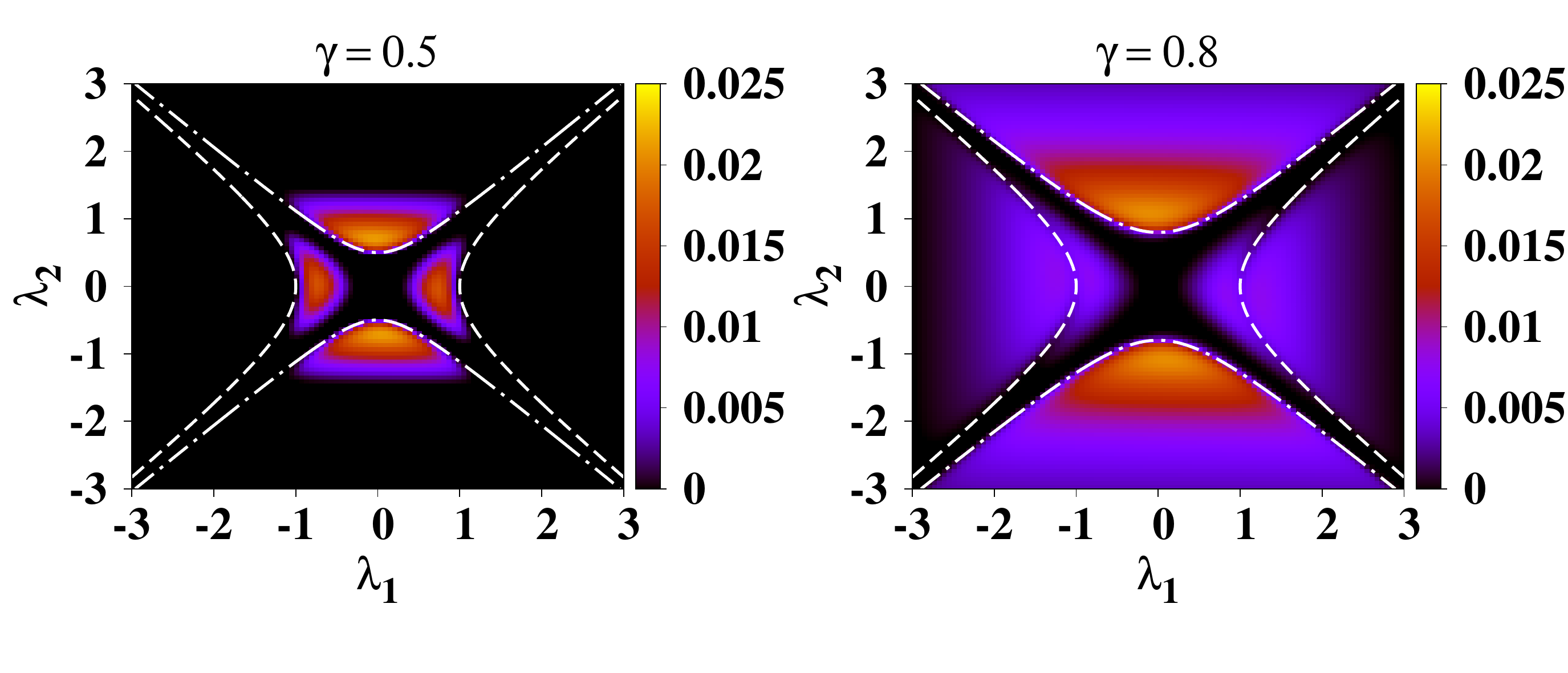}
 \caption{(Color online) Ergodicity score, $\eta_{\mathcal{S}}^\mathcal{D}$, corresponding to QD, as a function of $\lambda_1$ and $\lambda_2$ for 
 $\gamma=0.5$ and $\gamma=0.8$.
 The phase boundaries are same as in Fig. \ref{mono}. Here,  $\lambda_1, \lambda_2$ are dimensionless, and  $\eta_{\mathcal{S}}^{\mathcal{D}}$ is in bits.}
 \label{ergophase}
\end{figure}

In the case of bipartite quantum correlations, we consider the time average of the quantity at ``large'' time, $t_{L}$. 
The definitions of large time may vary depending
on the situation in hand. In general, we call a time instant, $t_L$, to be ``large'' if any one of the following scenarios occur. 
\begin{enumerate}
\item[\textbf{(a)}] $\mathcal{Q}$ saturates to $\mathcal{Q}_c$ for $t\geq t_L$, and remains constant at $\mathcal{Q}=\mathcal{Q}_c$ for 
$t\geq t_L$.

\item[\textbf{(b)}] $\mathcal{Q}$ oscillates for $t\geq t_L$, such that $\delta_{\mathcal{Q}}\leq \delta$.  Here, 
$\delta_{Q}$ is the amplitude of fluctuation in the values of $\mathcal{Q}$
for $t\geq t_L$, and $\delta$ is a small quantity whose value provides the required precision in determining $\mathcal{Q}$.

\item[\textbf{(c)}] For $t\geq t_L$, $\delta_{\mathcal{Q}}$ has a finite value, which remains constant in time. 
\end{enumerate}
\noindent Evidently, the time-average is not required in the case of \textbf{(a)} and \textbf{(b)}.

To determine ergodicity of the bipartite quantum correlations, as measured by LN and QD, we compute the value of 
$\eta_{\mathcal{S}}^\mathcal{L}$ and $\eta_{\mathcal{S}}^\mathcal{D}$, corresponding to LN and QD, respectively, for the points on the $(\lambda_1,\lambda_2)$ plane, with different values of $\gamma$. The initial CES at $t=0$ 
is chosen to be the one with $\beta J=100$. The values of both LN and QD tend to show the behavior described in \textbf{(c)}
for $J t\rightarrow J t_L$, which we found to be $\sim100\pi$. To determine the time averaged values of LN and QD, which depend on the choice of the values of the system parameters, we consider an interval of $20\pi$, starting from $J t=100\pi$. We analyse the ergodicity properties of LN and QD via three specific cases, as follows. 

In the first case, (\textbf{Case I.}) we take $\lambda_1=0.0,\lambda_2=0.15$, 
which is a point in the AFM region. In the left panel of Fig. \ref{ergoinst}, the time-averaged value of LN at large time, starting from a 
CES with $\beta J=100$ at $J t=0$, is represented by a dashed line, which is intersected by the graph of LN varying with $\beta$ (solid line). 
This, according to Eq. (\ref{ergo_score}), implies that LN is ergodic in this case. Similar conclusion about QD can be drawn, as 
depicted from the right panel of Fig. \ref{ergoinst}. However, QD does not always remain ergodic, as can be seen from \textbf{Case II.}
Here, we take a point in the DM phase, given by $\lambda_1=1,\lambda_2=2$, and see that the time-averaged value of LN at large time is zero
(left panel, Fig. \ref{ergoinst}), leading to 
ergodicity of LN. In contrast, the time-averaged QD at large time, depicted by the double-dotted line in the right panel of Fig. 
\ref{ergoinst}, does not coincide with QD of any CES for all $\beta J$, (solid line). Hence, QD is nonergodic in this case.

The above examples naturally leads to the question as to whether bipartite 
entanglement in the present model is always ergodic. To verify this, we 
perform extensive numerical search in the parameter space of $(\lambda_1,\lambda_2)$. We find that that bipartite entanglement, remains ergodic over the entire $(\lambda_1,\lambda_2)$ plane, up to our numerical accuracy (accurate up to the third decimal place). 
However, there are very small sets of values of $\lambda_1$ and $\lambda_2$, for which the status of ergodicity of LN remains inconclusive. One such 
instance is presented by a third case, \textbf{Case III}. Here, $\lambda_1=0.15$ and $\lambda_2=0.0$, representing a point in the AFM phase. The 
corresponding time-averaged value of LN is shown by dot-dashed line in the left panel of Fig. \ref{ergoinst}. We find that 
$\eta_{\mathcal{S}}^{\mathcal{L}}$, corresponding to LN, is zero up to the third decimal place -- the point to which we claim our 
data to be accurate. However, there is a possibility of obtaining non-zero values of $\eta_{\mathcal{S}}^{\mathcal{L}}$ 
with increased accuracy, which would imply that LN is nonergodic at $(\lambda_1=0.15,\lambda_2=0.0)$. Our numerical search suggests that 
the area of such regions on the $(\lambda_1,\lambda_2)$ plane is negligibly small (cf. \cite{dyn-ergo-xy-dual}).  From exclusive numerical simulations we possibly conclude that except for $\lambda_1,\lambda_2\approx 0$, bipartite entanglement is always ergodic irrespective of $\gamma$ and low values of $\beta$ of the initial state upto the numerical accuracy. 
In contrast, QD exhibits nonergodicity in the \textbf{Case III}. To investigate the ergodicity of QD over the $(\lambda_1,\lambda_2)$ plane, we 
compute $\eta_{\mathcal{S}}^{\mathcal{D}}$, corresponding to QD, as a function of $\lambda_1$ and $\lambda_2$. We find that 
the region of nonergodicity is small for small $\gamma$, and  grows over the $(\lambda_1,\lambda_2)$ plane, when the value of $\gamma$ is increased.
This can be understood from Fig. \ref{ergophase}, where 
the plots of the values of $\eta_{\mathcal{S}}^{\mathcal{D}}$ as function of $\lambda_1$ and $\lambda_2$ for different values of $\gamma$ are depicted.  
We have also plotted the zero-temperature QPT lines and the separable lines for comparison. 
Note that even for fairly high values of $\gamma$, QD in almost the entire AFM phase remains ergodic, while the nonergodicity in QD is 
most prominent in the DM phase near the AFM $\leftrightarrow$ DM QPT line. 

\begin{figure}
 \includegraphics[scale=0.3125]{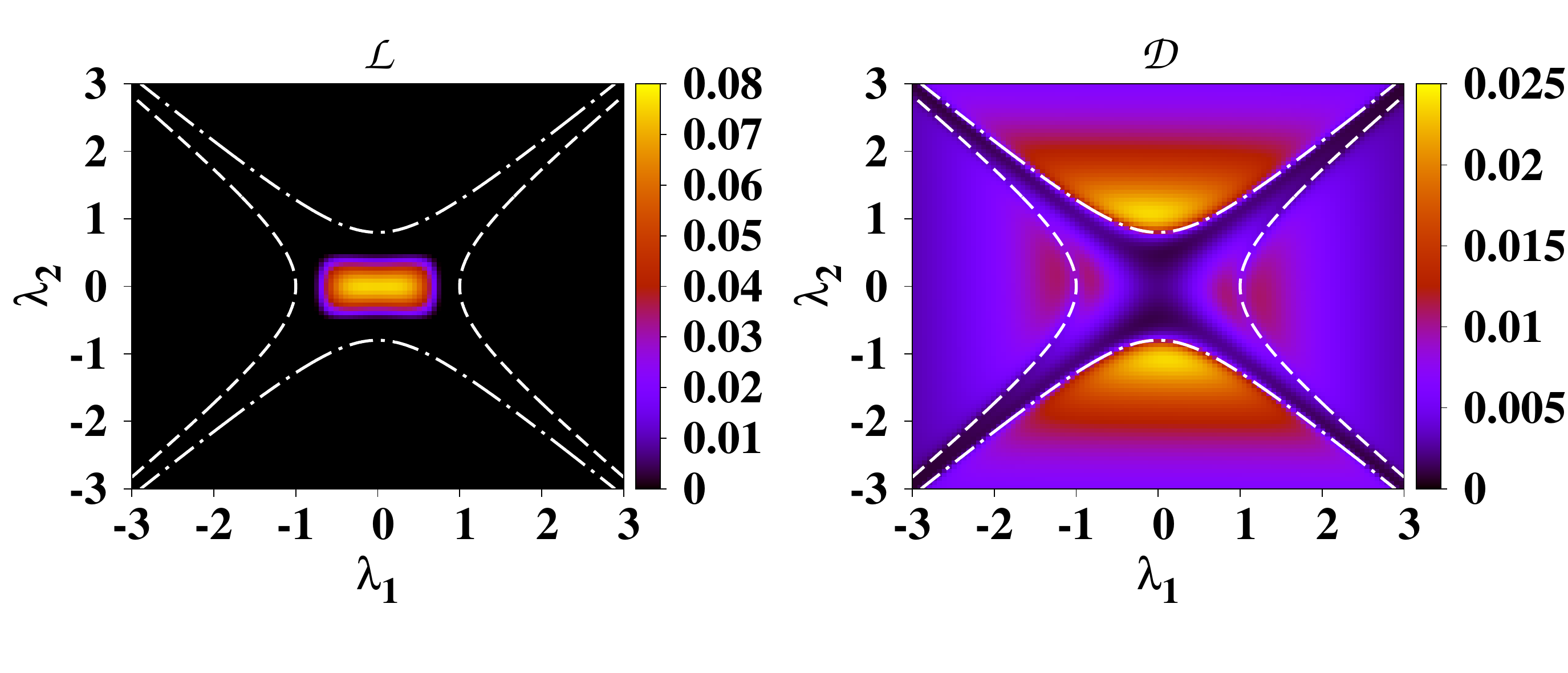}
 \caption{(Color online) Plot of the time-averaged LN (left panel) and QD (right panel) at large time as a function of $\lambda_1$ and $\lambda_2$, which are dimensionless.
 The phase boundaries are same as Fig. \ref{mono}. LN and QD are measured respectively in ebits and bits.
 }
 \label{larget}
\end{figure}

We conclude the discussion on ergodicity with a description of the variation of time-averaged LN and QD at large time  ($Jt\geq Jt_L$). 
Fig. \ref{larget} depicts the landscape of time-averaged values of LN and QD over the $(\lambda_1,\lambda_2)$-plane, where we have chosen $\gamma=0.8$
for discussion, and the initial state of the time evolution to be the CES at $\beta J=100$. It is clear from the figure that at $J t\geq J t_L$, LN
persists only in the AFM region, while it vanishes completely in the entire PM and DM phase. One must note here that the definition 
of ergodicity score in Eq. (\ref{ergo_score}), and the fact that entanglement may decrease with an increasing $\beta J$ imply that 
probability of finding a set of parameters, for which LN becomes non-ergodic, is higher in AFM phase where
the time-averaged LN at large $t$ has a non-zero value. This is in agreement with the \textbf{Case III} reported above, since 
the parameter values $(\lambda_1=0.15,\lambda_2=0.0)$ are in the region of the $(\lambda_1,\lambda_2)$ plane, where time-averaged value of LN at large time is high.

\begin{figure}
 \includegraphics[scale=0.3175]{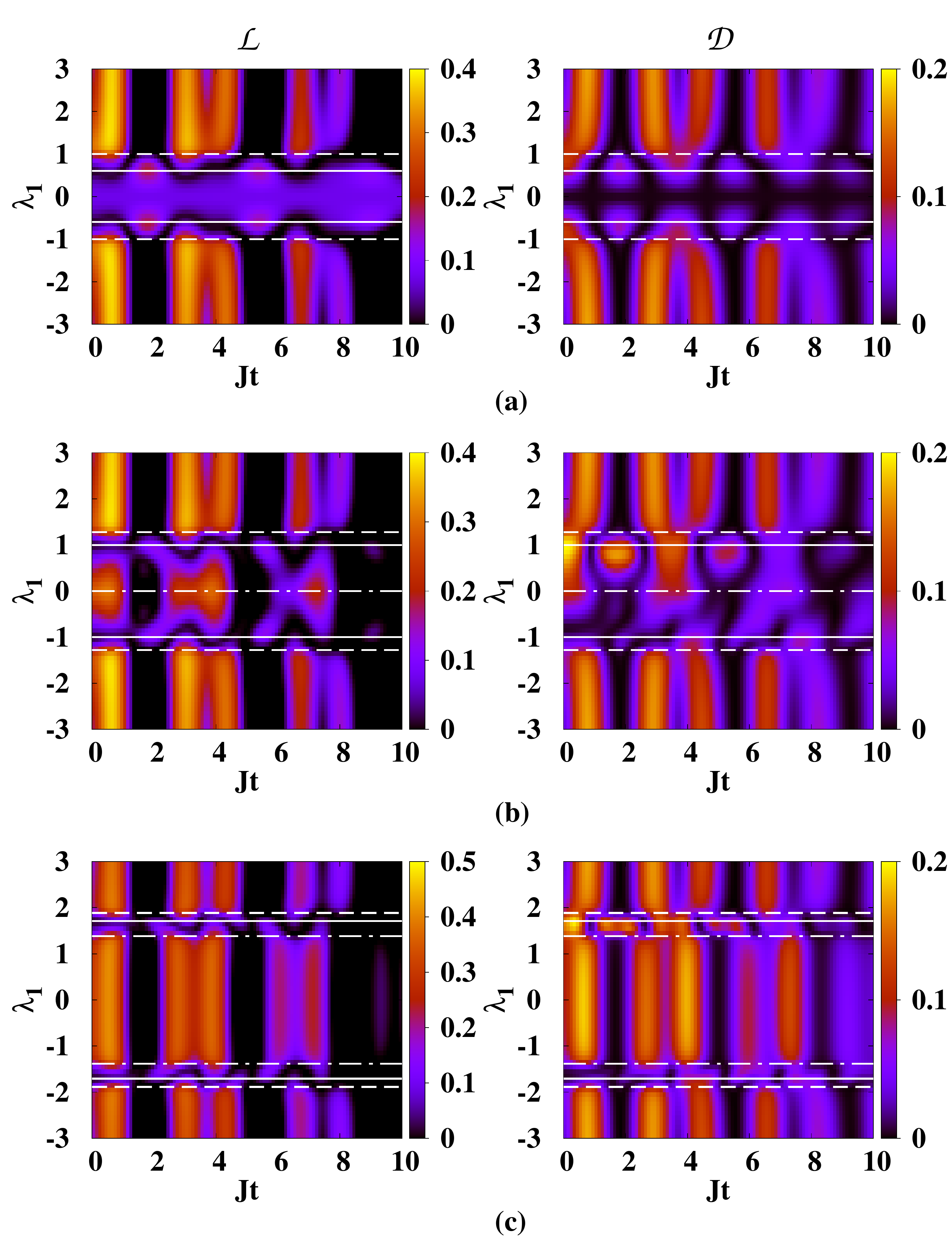}
 \caption{(Color online) Field-time landscape of  LN (left panels) and QD (right panels), where $\lambda_1$ is chosen to be 
 the varying field.
 Three different values of $\lambda_2$ have been chosen, viz. (a) $\lambda_2 = 0$, (b) $\lambda_2 = 0.8$, and (c) $\lambda_2 = 1.6$ with $\gamma = 0.8$.
 For comparison, we mark the different phases and the factorization line on the $\lambda_1$ axes (at $\beta\rightarrow\infty$), 
 indicated by the horizontal lines, same as Fig. \ref{mono}. 
 $J t$, $\lambda_1$ and $\lambda_2$ are dimensionless. LN and QD are respectively in ebits and bits.
 }
 \label{h2time}
\end{figure}

\begin{figure}[h]
 \includegraphics[scale=0.3175]{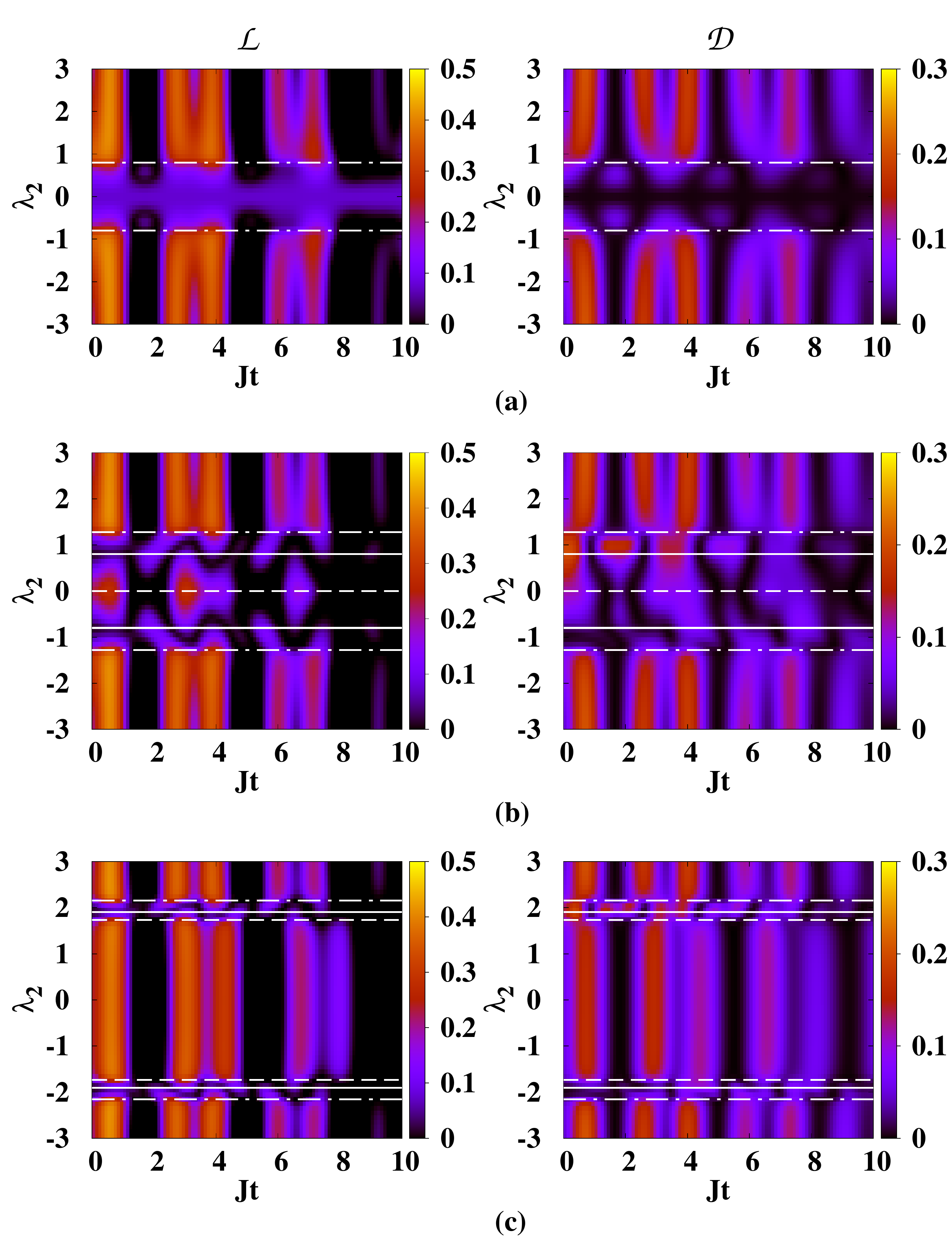}
 \caption{(Color online) Field-time landscape of  LN (left panels) and QD (right panels).
 (a) $\lambda_1 = 0$, (b) $\lambda_1 = 0.8$, and (c) $\lambda_1 = 1.6$. 
All other parameters and lines are same as Fig. \ref{h2time}.
}
 \label{h1time}
\end{figure}

\subsection{Dynamics at small time}

The question of ergodicity, of a physical quantity is important from the point of view of statistical mechanics. On the other hand, the information theoretic aspects demands the study of quantum correlation in the dynamics with small time. We fix the range to $0\leq J t\leq 4\pi$,  which is $25\%$ of the value of  $J t_L$.
Figs. \ref{h2time} and \ref{h1time} depicts the bird's-eyeview of the landscapes of LN and QD over the $(\lambda_{1(2)},Jt)$ plane, where $\lambda_{2(1)}$ 
is constant, and $t$ is in the range of small time. 
For typical fixed values of the set of systems parameters given by $(\lambda_1,\lambda_2)$, both LN and QD are found to collapse 
and revive non-periodically. 
It is clear from the Figs. \ref{h2time} and 
\ref{h1time} that the collapse of LN is more frequent than the collapse of QD at short time, although LN possesses much higher value.

\section{Discussions}
\label{conclude}

To summarize, we have considered a one-dimensional 
anisotropic XY chain of spin-$\frac{1}{2}$ spins, in the presence of a uniform and an alternating transverse field 
whose direction depends on
whether the lattice site is even or odd. 
The model, via a Jordan-Wigner transformation, can be mapped onto 
one-dimensional two-component Fermi gas defined on an optical lattice, constituted of two sublattices consisting of the even and the odd sites.
Although the analytical treatment of the model is similar to the well-known XY model, 
the system possesses new dimer phase apart from paramagnetic and anti-ferromagnetic phase.
We determine the singe-site magnetizations and two-site 
spin correlation functions corresponding to a nearest-neighbor spin pair in canonical equilibrium state and the time-evolved state of the 
model, and determine the nearest-neighbor density matrix. 
We study the static and dynamical characteristics of nearest neighbour entanglement quantified by LN
and by investigating their variations with 
relevant system parameters, temperature, and time. We determine the finite-size scaling exponents for the entanglement in the vicinity of the QPTs at zero temperature of the model. At finite temperature, we show that against increasing temperature,  
the bipartite entanglement is most fragile in the AFM phase, while being the most robust in the DM phase. We also demonstrate the occurrence 
of nonmonotonic variation of bipartite entanglement with temperature. We map the regions in different phases of the model on the plane 
of the chosen system parameter for which nonmonotonic variations of entanglement is found.
The trend of QD which is different from entanglement in the AFM phase, the region of nonmonotonicity grows with anisotropy, and covers almost the entire AFM phase when anisotropy
is high. However, the dimer phase remains completely free of such region for LN in the case of both high and low value of the 
anisotropy parameter has also been investigated and the measure found to be a tool for identifying phases present in this model.
 We find that when anisotropy in the system is low, 
nonmonotonicity for QD occurs mostly in the paramagnetic phase, while at high anisotropy, such regions shrink drastically. 

We also consider the dynamics of the bipartite quantum correlations, as measured by LN and QD.
We address the question of ergodicity of the bipartite correlations by looking into 
the ergodicity score corresponding to the chosen quantum correlation measure. We show that if canonical equilibrium state at a very low 
temperature is chosen to be a initial state of the evolution, up to our numerical accuracy, 
entanglement remains ergodic over the entire phase plane of the system. On the other hand, QD can be both ergodic as well as  nonergodic, 
suggesting an ``ergodic to nonergodic'' transition in the space of system parameters.
Benchmarking the time at which the dynamics of the quantum correlations equilibrates, 
we also define a range of short time, and discuss the short-time dynamics of LN and QD for the model in focus.  

\begin{acknowledgments}
The Authors thank D. K. Mukherjee, S. Bhattacharyya, and S. Singha Roy for fruitful discussions, and acknowledge computations performed at the cluster
computing facility of Harish-Chandra Research Institute.

\end{acknowledgments}

\appendix

\section{Diagonalization of the \texorpdfstring{$p^{th}$}{} subspace}
\label{ap:diagonalization}

The Hamiltonian $\hat{H}_p$ that acts on the $p^{th}$ subspace of dimension $16$ can be block-diagonalized by a choice of basis
$\{|\psi_i\rangle:1,\cdots,16\}$, given by
\begin{eqnarray}
|\psi_{1}\rangle &=& a_p^{\dagger}b_{p}^{\dagger}|0\rangle,\nonumber\\
|\psi_{2}\rangle &=& a_{-p}^{\dagger} b_{-p}^{\dagger}|0\rangle,  
\label{psub_1}
\end{eqnarray}

\begin{eqnarray}
|\psi_3\rangle &=& a_p^{\dagger}|0\rangle,\nonumber\\
|\psi_4\rangle &=&  b_{p}^{\dagger}|0\rangle,\nonumber\\
|\psi_5\rangle &=& a_p^{\dagger} a_{-p}^{\dagger} b_{p}^{\dagger}|0\rangle,\nonumber\\
|\psi_{6}\rangle &=& a_p^{\dagger} b_{p}^{\dagger}b_{-p}^{\dagger}|0\rangle, 
\label{psub_2}
\end{eqnarray}
\begin{eqnarray}
|\psi_{7}\rangle &=& a_{-p}^{\dagger}|0\rangle,\nonumber\\
|\psi_{8}\rangle &=& b_{-p}^{\dagger}|0\rangle\nonumber\\
|\psi_{9}\rangle &=& a_p^{\dagger} a_{-p}^{\dagger} b_{-p}^{\dagger}|0\rangle,\nonumber\\
|\psi_{10}\rangle &=& a_{-p}^{\dagger} b_{p}^{\dagger}b_{-p}^{\dagger}|0\rangle,
\label{psub_3}
\end{eqnarray}

\begin{eqnarray}
|\psi_{11}\rangle &=& a_p^{\dagger} b_{-p}^{\dagger}|0\rangle\nonumber\\
|\psi_{12}\rangle &=& a_{-p}^{\dagger} b_{p}^{\dagger}|0\rangle\nonumber\\
|\psi_{13}\rangle &=& a_p^{\dagger} a_{-p}^{\dagger}|0\rangle,\nonumber\\
|\psi_{14}\rangle &=& b_p^{\dagger} b_{-p}^{\dagger}|0\rangle,\nonumber\\
|\psi_{15}\rangle &=& a_p^{\dagger}a_{-p}^{\dagger} b_{p}^{\dagger}b_{-p}^{\dagger}|0\rangle,\nonumber\\
|\psi_{16}\rangle &=& |0\rangle,
\label{psub_4}
\end{eqnarray}
where $|0\rangle$ denotes the vacuum state. 
Note that the above sets of basis block-diagonalizes $\hat{H}_p$ into four blocks of dimensions $2$, $4$, $4$, and $6$, such that 
$\hat{H}_p=\bigoplus_{k=1}^{4}\hat{H}_p^k$, which explains the above distribution of the basis vectors into four groups, 
given by (\ref{psub_1})-(\ref{psub_4}). Using the form of 
$\hat{H}_p$ (Eq. (\ref{hp})), and Eqs. (\ref{psub_1})-(\ref{psub_4}), $\hat{H}_p^1$ is found to be a null matrix of dimension $2$, while
$\hat{H}_p^2=\hat{H}_p^3$, with 
\begin{eqnarray}
\hat{H}_p^2&=&
\left[
\begin{array}{cccc}
-h_1-h_2 & J\cos{\phi_p} & -iJ\gamma \sin{\phi_p} & 0 \\
 J\cos{\phi_p} & -h_1+h_2 & 0 & -iJ\gamma \sin{\phi_p}\\
iJ\gamma\sin{\phi_p} & 0 & h_1-h_2 & -J\cos\phi_p \\
 0 & iJ\gamma\sin{\phi_p} & -J\cos\phi_p & h_1+h_2  
\end{array}
\right],\nonumber \\
\label{Hp2}
\end{eqnarray}
and 
\begin{widetext}
\begin{eqnarray}
\hat{H}_p^4&=& 
\left[
\begin{array}{cccccc}
-2h_1 & iJ\gamma \sin{\phi_p} & -iJ\gamma \sin{\phi_p} & 0 & 0 &0 \\
 -iJ\gamma\sin{\phi_p} & 0 & 0 & J\cos{\phi_p} & J \cos{\phi_p} & -iJ\gamma \sin{\phi_p}\\
iJ\gamma\sin{\phi_p} & 0 & 0 & -J\cos{\phi_p} & -J \cos{\phi_p} & iJ\gamma \sin{\phi_p}\\
0 & J\cos{\phi_p} & -J\cos{\phi_p} & -2h_2 & 0 & 0 \\
0 & J\cos{\phi_p} & -J\cos{\phi_p} & 0 & 2h_2 & 0 \\
0 & iJ\gamma \sin{\phi_p} & -iJ\gamma \sin{\phi_p} & 0 & 0 & 2h_1
\end{array} 
\right].
\label{Hp4}
\end{eqnarray}
\end{widetext}
Hence, diagonalization of the $p^{th}$ subspace of dimension $16$ reduces to the diagonalization of the 
irreducible operators $\{\hat{H}_p^k$, $k=1,2,3,4\}$. \textcolor{red}{Note that $\hat{H}_p^2$ and $\hat{H}_p^3$ provide four distinct eigenvalues in the spectrum of $H_p$, each of which is two-fold degenerate. These four eigenvalues are given by $\pm \omega_2^{\pm}(p)$ where $\omega_2^{\pm}(p) = \sqrt{x(p) \pm 2\sqrt{y(p)}}$. Here,  $x(p) = \lambda_1^2 + \lambda_2^2 + \cos^2{\phi_p} + \gamma^2\sin^2{\phi_p}$, and  $y(p) =\lambda_1^2(\lambda_2^2+\cos^2{\phi_p}) + \gamma^2\lambda_2^2\sin^2{\phi_p}$, where $\lambda_{1(2)}=h_{1(2)}/J$. Two of the six eigenvalues of $\hat{H}_p^4$ are zero, while the other four eigenvalues are given by $\pm \omega_4^{\pm}(p)$, where $\omega_4^{\pm}(p) =4\sqrt{x(p) \pm \sqrt{{x(p)}^2 -4 y(p)}}$.  Clearly, $- \omega_2^{+}(p)$ and $- \omega_4^{+}(p)$ are the minimum eigenvalues of  $\hat{H}_p^2$ and $\hat{H}_p^4$, respectively. It can also be checked that $- \omega_4^{+}(p)\leq -\omega_2^{+}(p)$ irrespective of the value of $p$. The ground state energy per site is obtained by $E_0=-\frac{1}{2\pi}\int_0^{\pi/2}\omega_4^{+}(p) dp$.}

\section{Measures of quantum correlations}
\label{ap:qc}

We now briefly discuss two specific measures, namely, logarithmic negativity and quantum discord, belonging to 
entanglement-separability and quantum information theoretic paradigm, respectively. 

\noindent\textbf{\textit{Negativity and logarithmic negativity.}}  
The negativity \cite{neg_group}, \({\cal N}(\rho_{AB})\), for a bipartite state \(\rho_{AB}\), is  the absolute value of 
the sum of all the negative eigenvalues of \(\rho_{AB}^{T_{A}}\), and is given by 
 \begin{equation}
  {\cal N}(\rho_{AB})=\frac{\|\rho_{AB}^{T_A}\|_1-1}{2},
  \label{eq:negativity}
 \end{equation}
where \(\rho_{AB}^{T_{A}}\) is obtained from $\rho_{AB}$ by
performing the partial transposition with respect to the subsystem \(A\) \cite{neg_part_group}. 
Here, $\|\rho\|_1 \equiv \mbox{tr}\sqrt{\rho^\dag \rho}$ is the trace-norm of the matrix $\rho$.
The logarithmic negativity (LN) \cite{neg_group}, $\mathcal{L}(\rho_{AB})$, defined in terms of negativity, is given by 
\begin{equation}
\mathcal{L}(\rho_{AB}) = \log_2 [2 {\cal N}(\rho_{AB}) + 1].
\label{eq:LN}
\end{equation}

\noindent\textbf{\textit{Quantum Discord.}} 
Quantum discord \cite{disc_group} of a bipartite quantum state $\rho_{AB}$ is defined as the difference between the 
total correlation \cite{total_corr}, quantified by the quantum mutual information, and the classical correlation present in the system.  
The quantum mutual information is given by 
\begin{equation}
\mathcal{I}(\rho_{AB})= S(\rho_A)+ S(\rho_B)- S(\rho_{AB}),
\label{mutual_info}
\end{equation}
where $\rho_{A(B)}$ are the local density matrices of $\rho_{AB}$, obtained as $\rho_{A(B)}=\mbox{tr}_{B(A)}\left[\rho_{AB}\right]$, 
and $S(\varrho)= - \mbox{tr} (\varrho \log_2 \varrho)$ is the von Neumann entropy. 
The classical correlation of the state $\rho_{AB}$ is defined as 
\begin{equation}
 {\cal J}(\rho_{AB}) = S(\rho_B) - S(\rho_{B|A}),
 \label{class_corr}
\end{equation}
where $S(\rho_{B|A})$, the conditional entropy, is given by
\begin{equation}
 S(\rho_{B|A}) = \min_{\{P_i\}} \sum_i p_i S(\rho_{B|i}).
\end{equation} 
Here, $S(\rho_{B|A})$ is conditioned over the measurements performed on \(A\) via a rank-one projective measurements \(\{P_i\}\),
which produces the states \(\rho_{B|i} = \frac{1}{p_i} \mbox{tr}_A[(P_i \otimes \mathbb{I}_B) \rho_{AB} (P_i \otimes \mathbb{I}_B)]\), 
with probabilities \(p_i = \mbox{tr}[(P_i \otimes \mathbb{I}_B) \rho_{AB} (P_i \otimes \mathbb{I}_B)]\),  
and \(\mathbb{I}_B\) is the identity operator in the Hilbert space of \(B\). From Eqs. (\ref{mutual_info}) and 
(\ref{class_corr}), quantum discord can be obtained as  
\begin{equation}
{\cal D}(\rho_{AB})= {\cal I}(\rho_{AB}) - {\cal J}(\rho_{AB}).
\label{qdiscord}
\end{equation} 

\section{Two-site spin correlators}
\label{ap:corr_fun}
Similar to the Hamiltonian $\hat{H}_p$, the two-site spin correlator operator $\hat{c}^{\alpha\alpha}$, $\alpha=x,y$,
can be obtained as $\hat{c}^{\alpha\alpha}=\frac{2}{N}\sum_{p=1}^{N/4}\hat{c}^{\alpha\alpha}_p$, where in the $p^{th}$ subspace, 
$\hat{c}^{\alpha\alpha}_p$ is block-diagonalizable in the same basis as given in Appendix \ref{ap:diagonalization}. For example, one can obtain 
$\hat{c}^{xx}=\frac{2}{N}\sum_{i=1}^{N/2}\sigma_{2i}^x\sigma_{2i+1}^x$ corresponding to an ``even-odd'' pair of spins in the momentum space, 
such that 
\begin{eqnarray}
 \hat{c}^{xx}_p&=&e^{i\phi_p}\big( {b_{-p}}^{\dagger}a_{-p} -{a_p}^{\dagger}b_{-p}^{\dagger} +  {a_{p}}^{\dagger}b_{p} + {a_{-p}}b_{p}\big) \nonumber \\
&&+  e^{-i\phi_p}\big( {b_{p}}^{\dagger}a_{p} -{a_{-p}}^{\dagger}b_{p}^{\dagger} +  {a_{-p}}^{\dagger}b_{-p} + {a_{p}}b_{-p}\big).\nonumber \\
\end{eqnarray}
In the basis given in Appendix \ref{ap:diagonalization}, one can write $\hat{c}^{\alpha\alpha}_p=\bigoplus_{k=1}^4\hat{c}^{xx,k}_p$, 
where $\hat{c}^{xx,1}_p$ is a 
null matrix of dimension $2$, and $\hat{c}^{xx,2}_p$, $\hat{c}^{xx,3}_p$, and $\hat{c}^{xx,4}_p$ are given by  
\begin{widetext}
 \begin{eqnarray}
 \hat{c}^{xx,2}_p&=&
\left[
\begin{array}{cccc}
0 & e^{i\phi_p} & -e^{i\phi_p} & 0  \\
    e^{-i\phi_p} & 0 & 0 & e^{-i\phi_p}  \\
    -e^{-i\phi_p} & 0 & 0 & -e^{-i\phi_p} \\
    0 & e^{i\phi_p} & -e^{i\phi_p} & 0 
\end{array}
\right],\,\,
\hat{c}^{xx,3}_p= 
\left[
\begin{array}{cccc}
0 & e^{-i\phi_p} & e^{-i\phi_p} & 0  \\
    e^{i\phi_p} & 0 & 0 & -e^{i\phi_p}  \\
    e^{i\phi_p} & 0 & 0 & -e^{i\phi_p} \\
    0 & -e^{-i\phi_p} & -e^{-i\phi_p} & 0 
\end{array}
\right],\nonumber\\
  \hat{c}^{xx,4}_p&=&
  \left[ 
  \begin{array}{cccccc}
    0 & -e^{-i\phi_p} & -e^{i\phi_p} & 0 & 0 & 0 \\
    -e^{i\phi_p} & 0 & 0 & e^{i\phi_p} & e^{i\phi_p} & -e^{i\phi_p} \\
    -e^{-i\phi_p} & 0 & 0 & -e^{-i\phi_p} & -e^{-i\phi_p} & -e^{-i\phi_p} \\
    0 & e^{-i\phi_p} & -e^{i\phi_p} & 0 & 0 & 0 \\
    0 & e^{-i\phi_p} & -e^{i\phi_p} & 0 & 0 & 0 \\
    0 & -e^{-i\phi_p} & -e^{i\phi_p} & 0 & 0 & 0 \\
    \end{array}
   \right].  
 \end{eqnarray}
\end{widetext}
Similar calculation for $\hat{c}^{yy}$ leads to $\hat{c}^{yy,1}_p=c^{xx,1}_p$, and 
\begin{widetext}
\begin{eqnarray}
    \hat{c}^{yy,2} &=&
    \left[ 
    \begin{array}{cccc}
    0 & e^{i\phi_p} & e^{i\phi_p} & 0  \\
    e^{-i\phi_p} & 0 & 0 & -e^{-i\phi_p}  \\
    e^{-i\phi_p} & 0 & 0 & -e^{-i\phi_p} \\
    0 & -e^{i\phi_p} & -e^{i\phi_p} & 0 
    \end{array}
    \right],\,\,
    \hat{c}^{yy,3}_p = 
    \left[
    \begin{array}{cccc}
    0 & e^{-i\phi_p} & -e^{-i\phi_p} & 0  \\
    e^{i\phi_p} & 0 & 0 & e^{i\phi_p}  \\
    -e^{i\phi_p} & 0 & 0 & -e^{i\phi_p} \\
    0 & e^{-i\phi_p} & -e^{-i\phi_p} & 0 
    \end{array}
    \right],\nonumber\\
    \hat{c}^{yy,4}_p &=& 
    \left[
    \begin{array}{cccccc}
    0 & e^{-i\phi_p} & e^{i\phi_p} & 0 & 0 & 0 \\
    e^{i\phi_p} & 0 & 0 & e^{i\phi_p} & e^{i\phi_p} & -e^{i\phi_p} \\
    e^{-i\phi_p} & 0 & 0 & -e^{-i\phi_p} & -e^{-i\phi_p} & e^{-i\phi_p} \\
    0 & e^{-i\phi_p} & -e^{i\phi_p} & 0 & 0 & 0 \\
    0 & e^{-i\phi_p} & -e^{i\phi_p} & 0 & 0 & 0 \\
    0 & e^{-i\phi_p} & e^{i\phi_p} & 0 & 0 & 0 \\
    \end{array}
    \right]
\end{eqnarray}
\end{widetext}
Moreover, in the case of time-evolution, the operators $\hat{c}^{xy}$ and $\hat{c}^{yx}$ are given by
\begin{widetext}
\begin{eqnarray}
    \hat{c}^{xy,2} &=&-i
    \left[ 
    \begin{array}{cccc}
    0 & e^{-i\phi_p} & -e^{-i\phi_p} & 0  \\
    -e^{i\phi_p} & 0 & 0 & e^{i\phi_p}  \\
    e^{i\phi_p} & 0 & 0 & -e^{i\phi_p} \\
    0 & -e^{-i\phi_p} & e^{-i\phi_p} & 0 
    \end{array}
    \right],\,\,
    \hat{c}^{xy,3}_p = -i
    \left[
    \begin{array}{cccc}
    0 & e^{-i\phi_p} & -e^{-i\phi_p} & 0  \\
    e^{i\phi_p} & 0 & 0 & e^{i\phi_p}  \\
    -e^{i\phi_p} & 0 & 0 & -e^{i\phi_p} \\
    0 & e^{-i\phi_p} & -e^{-i\phi_p} & 0 
    \end{array}
    \right],\nonumber\\
    \hat{c}^{xy,4}_p &=& -i
    \left[
    \begin{array}{cccccc}
     0 & e^{-i\phi_p} & e^{i\phi_p} & 0 & 0 & 0 \\
    -e^{i\phi_p} & 0 & 0 & -e^{i\phi_p} & e^{i\phi_p} & e^{i\phi_p} \\
    -e^{-i\phi_p} & 0 & 0 & e^{-i\phi_p} & -e^{-i\phi_p} & e^{-i\phi_p} \\
    0 & e^{-i\phi_p} & -e^{i\phi_p} & 0 & 0 & 0 \\
    0 & -e^{-i\phi_p} & e^{i\phi_p} & 0 & 0 & 0 \\
    0 & -e^{-i\phi_p} & -e^{i\phi_p} & 0 & 0 & 0 \\
    \end{array}
    \right]
\end{eqnarray}
and
\begin{eqnarray}
    \hat{c}^{yx,2} &=&-i
    \left[ 
    \begin{array}{cccc}
    0 & -e^{-i\phi_p} & -e^{-i\phi_p} & 0  \\
    e^{i\phi_p} & 0 & 0 & e^{i\phi_p}  \\
    e^{i\phi_p} & 0 & 0 & e^{i\phi_p} \\
    0 & -e^{-i\phi_p} & -e^{-i\phi_p} & 0 
    \end{array}
    \right],\,\,
    \hat{c}^{yx,3}_p = -i
    \left[
    \begin{array}{cccc}
    0 & -e^{i\phi_p} & e^{i\phi_p} & 0  \\
    e^{-i\phi_p} & 0 & 0 & -e^{-i\phi_p}  \\
    -e^{-i\phi_p} & 0 & 0 & e^{-i\phi_p} \\
    0 & e^{i\phi_p} & -e^{i\phi_p} & 0 
    \end{array}
    \right],\nonumber\\
    \hat{c}^{yx,4}_p &=& -i
    \left[
    \begin{array}{cccccc}
    0 & e^{-i\phi_p} & e^{i\phi_p} & 0 & 0 & 0 \\
    -e^{i\phi_p} & 0 & 0 & e^{i\phi_p} & -e^{i\phi_p} & e^{i\phi_p} \\
    -e^{-i\phi_p} & 0 & 0 & -e^{-i\phi_p} & e^{-i\phi_p} & e^{-i\phi_p} \\
    0 & -e^{-i\phi_p} & e^{i\phi_p} & 0 & 0 & 0 \\
    0 & e^{-i\phi_p} & -e^{i\phi_p} & 0 & 0 & 0 \\
    0 & -e^{-i\phi_p} & -e^{i\phi_p} & 0 & 0 & 0 \\
    \end{array}
    \right],
\end{eqnarray}
\end{widetext}
with $\hat{c}^{xy,1}$ and $\hat{c}^{yx,1}$ being $2\times2$ null matrices.

\begin{figure*}
 \includegraphics[width=0.85\textwidth]{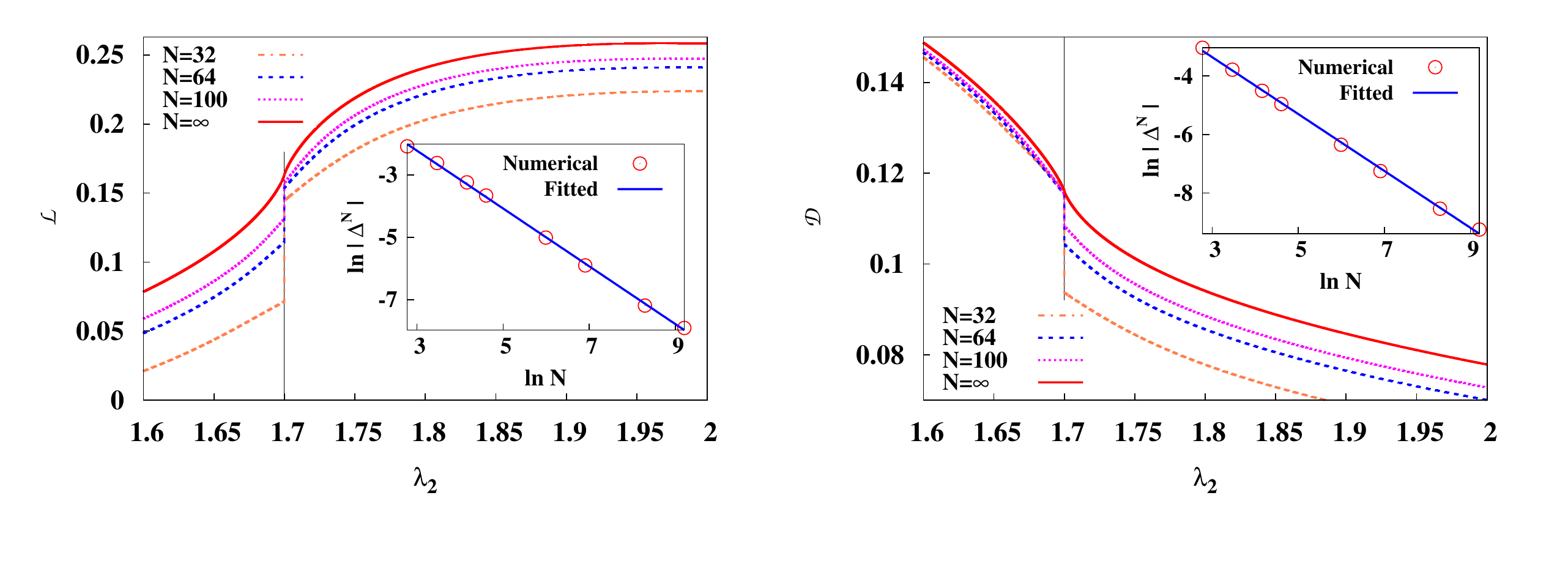}
 \caption{(Color online.) The figure in the left (right) panel depicts the variation of 
 $\mathcal{L}$ ($\mathcal{D}$) with $\lambda_2$ across the AFM 
 $\leftrightarrow$ DM QPT for different values of $N$, with $\lambda_1=1.5$. 
 At the QPT point, the quantum correlations 
 exhibit a finite jump in magnitude, given by $\Delta^N$. 
 (Insets) Corresponding variations of 
 $\ln\left|\Delta^N\right|$ (both numerical data and fitted line) as a function of $\ln N$. All the quantities plotted are dimensionless, 
 except LN which is in ebits and QD, that is in bits.}
 \label{scale-dm}
\end{figure*}

\section{Magnetization and correlation functions in CES}

For the nearest-neighbour reduced density matrix at $t=0$, 
one needs to determine the single-site magnetizations, $m^z_e$ and $m_o^z$, and the diagonal elements of 
the correlation tensor, $c^{\alpha\alpha}$, of $\rho_{eq}(t=0)$. In order to do so, we exploit the fact that 
the Hilbert space of the Hamiltonian (Eq. (\ref{ham_spin})) can be decomposed into non-interacting subspaces in the momentum space. 
In the $p^{th}$ such subspace of the momentum space, the CES can be written as $\rho_{eq}^p=\exp(-\beta\hat{H}_p(t=0))/Z_p$, where 
$Z_p=\mbox{Tr}[\exp(-\beta\hat{H}_p(t=0))]$ is the partition function in that momentum subspace. 
Using the form of $\rho^p_{eq}$, equilibrium expectation value of an operator $\hat{\mathcal{O}}_p$ can be obtained as 
\begin{eqnarray}
 \langle\hat{\mathcal{O}}\rangle=\frac{2}{N}\sum_{p=1}^{N/4}\mbox{Tr}[\hat{\mathcal{O}}_p\rho^p_{eq}]/\mbox{Tr}[\rho^p_{eq}].
 \label{exp_eql}
\end{eqnarray}
From the transformation scheme described in Sec. \ref{model},  the transverse magnetization operator in momentum space 
for an odd (even) site can be calculated as  
$\hat{m}^z_p=2(c_p^\dagger c_p+c_{-p}^\dagger c_{-p}-1)$, where $c\equiv a$ (odd site) or $c\equiv b$ (even site). 
We find that, similar to $\hat{H}_p$, the two-site correlator operators $\hat{c}^{\alpha\alpha}$, $\alpha=x,y$,
can be written as $\hat{c}^{\alpha\alpha}=\frac{2}{N}\sum_{p=1}^{N/4}\hat{c}^{\alpha\alpha}_p$, where 
$\hat{c}^{\alpha\alpha}_p$ can be expanded in the same basis 
as described in Appendix \ref{ap:diagonalization}. The forms of the operators $\hat{c}^{\alpha\alpha}_p$, in the momentum space, are given in 
Appendix \ref{ap:corr_fun}. Unlike $\hat{c}^{xx}$ and $\hat{c}^{yy}$, $\hat{c}^{zz}$ can not be obtained directly due to the presence of 
the four-fermionic terms in its expansion, but its expectation value, $c^{zz}$, can be obtained from the relation 
\begin{eqnarray}
c^{zz}=m^z_om^z_e-c^{xx}c^{yy},
\end{eqnarray}
for the thermal state including the zero-temperature state.
Here, we denote the expectation values of the respective operators by the same symbol without the hat. 
Note that in the thermodynamic limit $N\rightarrow\infty$, the sum in Eq. (\ref{exp_eql}) is replaced by an integral with proper limit
in the reduced Brillouin zone, such that Eq. (\ref{exp_eql}) reads 
\begin{eqnarray}
 \langle\hat{\mathcal{O}}\rangle=\frac{1}{\pi}\int_{0}^{\frac{\pi}{2}}
 \mbox{Tr}[\hat{\mathcal{O}}_p\rho^p_{eq}]/\mbox{Tr}[\rho^p_{eq}]d\phi_p.
 \label{thermo_lim}
\end{eqnarray}

\begin{table}
\begin{tabular}{|c|}
   \hline 
   Tuning parameter: $\lambda_2$\\
   \hline 
   \begin{tabular}{c|c|c}
      $\lambda_1$ & LN & QD\\
      \hline
      $0.0$ & \begin{tabular}{c}
                 $\tilde{\nu}_2=0.992 \pm 0.010$\\
                 $\ln\tilde{\alpha}_2=1.357  \pm  0.065$ \\
              \end{tabular}
            & \begin{tabular}{c}
                 $\tilde{\nu}_2=0.893 \pm 0.018$\\
                 $\ln\tilde{\alpha}_2=-1.453\pm 0.132$ \\
              \end{tabular}\\
      \hline
      $1.5$ & \begin{tabular}{c}
                 $\tilde{\nu}_2=0.926 \pm 0.010$\\
                 $\ln\tilde{\alpha}_2=0.559 \pm 0.063$ \\
              \end{tabular}
            & \begin{tabular}{c}
                 $\tilde{\nu}_2=0.972 \pm 0.015$\\
                 $\ln\tilde{\alpha}_2=-0.443 \pm  0.092$ \\
              \end{tabular}\\
   \end{tabular}\\
   \hline
\end{tabular}

\vspace{0.5cm}

\begin{tabular}{|c|}
   \hline 
   Tuning parameter: $\lambda_1$\\
   \hline 
   \begin{tabular}{c|c|c}
      $\lambda_2$ & LN & QD\\
      \hline
      $1.5$ & \begin{tabular}{c}
                 $\tilde{\nu}_1=0.942 \pm 0.009$\\
                 $\ln\tilde{\alpha}_1=0.721 \pm 0.054$ \\
              \end{tabular}
            & \begin{tabular}{c}
                 $\tilde{\nu}_1=0.988 \pm 0.013$\\
                 $\ln\tilde{\alpha}_1=-0.286  \pm 0.082$ \\
              \end{tabular}\\
   \end{tabular}\\
   \hline
\end{tabular}
\caption{The fitting parameters corresponding to the finite jumps of LN and QD at the AFM $\leftrightarrow$ DM transition point, 
arising out of the approximations used in the analysis. For all the computations, $\gamma=0.8$.}
\label{expo_afmdm_jump}
\end{table}

\section{AFM to DM transition}
\label{afmdm}
While investigating the AFM $\leftrightarrow$ DM QPT, one may try to determine the zero-temperature canonical equilibrium state $\rho_{eo}$
corresponding to a nearest-neighbour even-odd spin pair, by using the methodology discussed in Sec. \ref{diagonalization}. 
However, due to the approximations in the calculation, the variations of LN and QD exhibit finite jumps at the QPT point for fixed finite values 
of the system size $N$. This imposes a restriction in analyzing the finite size scaling behaviour using the usual procedure as discussed 
in the case of the AFM $\leftrightarrow$ DM. To understand this feature of the approximations properly, 
let us denote the value of the quantum correlation measure (which, in the present case, is either LN, or QD) 
by $Q_{-\delta}$, when $\lambda_{1(2)}=\lambda_{1(2)}^c-\delta$ with arbitrarily small $\delta (\rightarrow 0)$, while the 
same for $\lambda_{1(2)}=\lambda_{1(2)}^c+\delta$ is given by $Q_{+\delta}$. We find the trends of absolute value of the difference between 
$\mathcal{Q}_{\pm\delta}$ for 
a fixed value of $N$, denoted by $\Delta^N$ and it approaches zero with increasing $N$ as
\begin{eqnarray}
|\Delta^N|=\tilde{\alpha}_{1(2)}N^{-\tilde{\nu}_{1(2)}},
 \label{gapscale}
\end{eqnarray}
where $\tilde{\alpha}_{1(2)}$ is a dimensionless constant. Note that the subscript ``$1(2)$" indicates the choice of $\lambda_1(\lambda_2)$ as the tuning parameter. 
Insets of Fig. \ref{scale-dm} depicts the variations of $\ln |\Delta^N|$ as a function of $\ln N$. Values of $\tilde{\alpha}_{1(2)}$ and $\tilde{\nu}_{1(2)}$ can be estimated by fitting the numerical data with Eq. (\ref{gapscale}). 
The values of $\tilde{\alpha}$ and $\tilde{\nu}$ for LN and QD  
are given in Table \ref{expo_afmdm_jump}, 
where the values of $\lambda_2(\lambda_1)$ are kept fixed at $\lambda_1=0$ and $1.5$ ($\lambda_2=1.5$). 
This analysis indicates that the approximations are too drastic to investigate the intricacies of the AFM $\leftrightarrow$ DM transitions 
in the model. However, as expected, the effect of the approximations tends to disappear with increasing $N$. 


\end{document}